\documentclass[12pt]{article}
\usepackage{epsf}
\usepackage{hyperref}
\usepackage{longtable}

\newcommand{\bmat}{\left(\begin{array}}
\newcommand{\emat}{\end{array}\right)}

\def\yzero{\smash{\hbox{$y\kern-4pt\raise1pt\hbox{${}^\circ$}$}}}

\def\beq{\begin{equation}}
\def\eeq{\end{equation}}
\def\beqa{\begin{eqnarray}}
\def\eeqa{\end{eqnarray}}

\def\-{\hphantom{-}}

\def\s2{\frac{1}{\sqrt2}}

\def\beq{\begin{equation}}
\def\eeq{\end{equation}}
\def\beqa{\begin{eqnarray}}
\def\eeqa{\end{eqnarray}}

\def\diag{{\rm diag \,}}
\def\IF{\relax{\rm I\kern-.18em F}}
\def\II{\relax{\rm I\kern-.18em I}}
\def\IP{\relax{\rm I\kern-.18em P}}
\def\IC{\relax\hbox{\kern.25em$\inbar\kern-.3em{\rm C}$}}
\def\IR{\relax{\rm I\kern-.18em R}}

\def\Dsl{\,\raise.15ex\hbox{/}\mkern-13.5mu D} 
\def\IZ{Z\kern-.4em  Z}
\def\id{{\bf 1}}
\def\Ad{\bf \rm Ad}

\newcommand{\drawsquare}[2]{\hbox{%
\rule{#2pt}{#1pt}\hskip-#2pt
\rule{#1pt}{#2pt}\hskip-#1pt
\rule[#1pt]{#1pt}{#2pt}}\rule[#1pt]{#2pt}{#2pt}\hskip-#2pt
\rule{#2pt}{#1pt}}

\newcommand{\fund}{\raisebox{-.5pt}{\drawsquare{6.5}{0.4}}}
\newcommand{\Ysymm}{\raisebox{-.5pt}{\drawsquare{6.5}{0.4}}\hskip-0.4pt%
        \raisebox{-.5pt}{\drawsquare{6.5}{0.4}}}
\newcommand{\Yasymm}{\raisebox{-3.5pt}{\drawsquare{6.5}{0.4}}\hskip-6.9pt%
        \raisebox{3pt}{\drawsquare{6.5}{0.4}}}
\newcommand{\antifund}{\overline{\fund}}
\newcommand{\bYasymm}{\overline{\Yasymm}}
\newcommand{\bYsymm}{\overline{\Ysymm}}

\catcode`\@=11   
\newdimen\@rotdimen
\newbox\@rotbox  

\def\@vspec#1{\special{ps:#1}}
\def\@rotstart#1{\@vspec{gsave currentpoint currentpoint translate
   #1 neg exch neg exch translate}}
\def\@rotfinish{\@vspec{currentpoint grestore moveto}}
\def\@rotr#1{\@rotdimen=\ht#1\advance\@rotdimen by\dp#1%
   \hbox to\@rotdimen{\hskip\ht#1\vbox to\wd#1{\@rotstart{90 rotate}%
   \box#1\vss}\hss}\@rotfinish}
\def\@rotl#1{\@rotdimen=\ht#1\advance\@rotdimen by\dp#1%
   \hbox to\@rotdimen{\vbox to\wd#1{\vskip\wd#1\@rotstart{270 rotate}%
   \box#1\vss}\hss}\@rotfinish}%
\def\@rotu#1{\@rotdimen=\ht#1\advance\@rotdimen by\dp#1%
   \hbox to\wd#1{\hskip\wd#1\vbox to\@rotdimen{\vskip\@rotdimen
   \@rotstart{-1 dup scale}\box#1\vss}\hss}\@rotfinish}%
\def\@rotf#1{\hbox to\wd#1{\hskip\wd#1\@rotstart{-1 1 scale}%
   \box#1\hss}\@rotfinish}%
\def\rotate{\@ifnextchar[{\@rotate}{\@rotate[l]}}
\def\@rotate[#1]#2{\setbox\@rotbox=\hbox{#2}\@nameuse{@rot#1}\@rotbox}
\def\hre#1#2{\href{http://arxiv.org/abs/#1/#2}{[ArXiv:#1/#2]}}

\catcode`\@=12

\topmargin
-1.5cm
\textwidth
15.5cm
\textheight
23.5cm
\oddsidemargin
0.7cm
\evensidemargin
1.2cm

\begin{document}

\makeatletter
\@addtoreset{equation}{section}
\makeatother
\renewcommand{\theequation}{\thesection.\arabic{equation}}
\pagestyle{empty}
\rightline{ IFT-UAM/CSIC-07-12}
\rightline{ CERN-PH-TH/2007-061}
\rightline{\tt hep-th/yymmnnn}
\vspace{0.1cm}
\begin{center}
\LARGE{ Instanton Induced Neutrino Majorana Masses\\   
in   CFT Orientifolds with MSSM-like spectra
 \\[10mm]}
\large{ L.E. Ib\'a\~nez$^{1}$ A.N. Schellekens$^2$
 and A. M. Uranga$^{3}$ \\[3mm]}
\footnotesize{
${}^{1}$ Departamento de F\'{\i}sica Te\'orica C-XI and  
Instituto de F\'{\i}sica Te\'orica  C-XVI,\\[-0.3em]
Universidad Aut\'onoma de Madrid,
Cantoblanco, 28049 Madrid, Spain \\[2mm] 
 $^2$ NIKHEF, Kruislaan 409, 1009DB Amsterdam,
The Netherlands;\\
IMAPP, Radboud Universiteit, Nijmegen, The Netherlands;\\
Instituto de Matem\'aticas y F\'\i sica Fundamental, CSIC, \\
Serrano 123, Madrid 28006, Spain.\\[2mm]
${}^{3}$ PH-TH Division, CERN
CH-1211 Geneva 23, Switzerland\\
(On leave from IFT, Madrid, Spain)\\
[4mm]}
\small{\bf Abstract} \\[5mm]
\end{center}
\begin{center}
\begin{minipage}[h]{17.0cm}
{\small Recently it has been shown that string instanton effects may give rise to  neutrino 
Majorana masses in certain   classes of semi-realistic string compactifications. 
In  this paper we make a systematic search for supersymmetric MSSM-like Type II Gepner orientifold constructions  admitting  boundary states associated with 
 instantons giving rise to  neutrino Majorana masses and other L- and/or B-violating operators. We analyze the zero mode structure of D-brane instantons on general type II orientifold compactifications, and show that only instantons with $O(1)$  symmetry can have just the two zero modes required to contribute to the 4d superpotential.
We however discuss how the addition of fluxes and/or possible non-perturbative extensions of the orientifold compactifications would allow also instantons with $Sp(2)$ and $U(1)$ symmetries  to generate such superpotentials. In the context of Gepner orientifolds with MSSM-like spectra, we find no models with $O(1)$ instantons with just the required zero modes to generate a neutrino mass superpotential. On the other hand we find a number of  models in one particular orientifold of the Gepner model $(2,4,22,22)$ with $Sp(2)$ instantons with a few  extra uncharged non-chiral zero modes which could be easily lifted by the mentioned effects.   A few more orientifold  examples are also found under less stringent constraints on  the zero modes. 
This class of $Sp(2)$ instantons have the interesting property that R-parity conservation is automatic and the flavour structure of the neutrino Majorana mass matrices has a simple factorized form.}
\end{minipage}
\end{center}
\newpage
\setcounter{page}{1}
\pagestyle{plain}
\renewcommand{\thefootnote}{\arabic{footnote}}
\setcounter{footnote}{0}


\section{Introduction}

String Theory, as the leading  candidate for a  unified theory of Particle Physics and  Gravity, 
should be able to describe all observed particle  phenomena. One the most valuable 
experimental pieces of information obtained in the last decade concerns
neutrino masses. Indeed the evidence from solar, atmospheric, reactor and accelerator 
experiments indicates that neutrinos are massive. The simplest explanation of the smallness of
neutrino masses is the see-saw mechanism \cite{seesaw}. The SM gauge symmetry allows for two types of
operators bilinear on the neutrinos (with dimension $\le 4$) :
\beq
{\cal L}_{\nu }\ =\ M_{ab}\nu_{R}^a\nu_R^b \ +\ h_{ab}\nu_R^a{\bar H} L^b
\label{bilineales}
\eeq
where $\nu_R$ is the right-handed neutrino, $L$ is the left-handed lepton doublet 
and ${\bar H}$ is the Higgs field. In supersymmetric theories, this term arise from a superpotential with the above structure, upon replacing fields by chiral superfields.
If $M_{ab}$  is large, the lightest neutrino eigenvalues have masses 
\beq 
M_{\nu } \ =\   <{\bar H}>^2 h^{T} M^{-1} h 
\label{seesaw}
\eeq
For $M\sim 10^{10}-10^{13}$ GeV and Dirac neutrino masses of order  charged lepton masses, 
the eigenvalues are consistent with experimental results. 

What is the structure of neutrinos and their masses in string theory? 
 In specific compactifications 
giving rise to the  MSSM spectra singlet fields corresponding to right-handed 
neutrinos $\nu _R$ generically appear. Dirac neutrino masses are also generically present 
but the required Majorana $\nu_R$ masses are absent. This is because 
most MSSM-like models constructed to date have extra $U(1)$ symmetries, under which the right-handed neutrinos are charged, which hence forbid such masses. 
In many models, such symmetries are associated to a $U(1)_{B-L}$ gauge boson beyond  the SM.
In order to argue for the existence of $\nu_R$ masses, 
string model builders have searched for non-renormalizable couplings 
of the type $(\nu_R \nu_R  {\bar N}_R {\bar N}_R)$ with 
extra singlets ${N}_R$. Once the latter fields get a vev,  $U(1)_{B-L}$ is broken  and a Majorana 
mass appears for the $\nu_R$. Although indeed such couplings (or similar ones with higher
dimensions) exist in some semi-realistic compactifications, such a solution to the 
neutrino mass problem in string theory has  two problems: 1) The typical $\nu_R$ 
masses so generated  tend to be too small due to the higher dimension of the 
involved operators and 2) The vevs for the $N_R$ fields breaks spontaneously 
R-parity so that dimension 4 operators potentially giving rise to
fast proton decay are generated.  This is in a nutshell the neutrino problem in string compactifications (see \cite{Giedt:2005vx} for a recent discussion in heterotic setups).

In \cite{iu} (see also \cite{bcw}) two of the present authors pointed out that 
there is a built-in mechanism in string theory which may naturally give rise to 
Majorana masses for right-handed neutrinos. It was pointed out that string theory
instantons may generate such masses through operators of the general form
\beq 
M_{string}\, e^{-U}\, \nu_R\nu_R \ \ .
\label{quebonito}
\eeq
Here $U$ is a linear combination of closed string moduli whose imaginary part  gets shifted
under a $U(1)_{B-L}$ gauge transformation in such a way that the operator is fully
gauge invariant. The exponential factor comes from the semi-classical contribution 
of a certain class of string instantons. This a pure stringy effect distinct from 
the familiar gauge instanton effects which give rise to couplings violating 
anomalous global symmetries like $(B+L)$ in the SM. Here also  $(B-L)$ (which  
is anomaly-free) is violated. 

This operator is generated due to existence of instanton fermionic zero modes 
which are charged under $(B-L)$ and couple to the $\nu_R$ chiral superfield. 
Although the effect can take place in different constructions,
the most intuitive description may be obtained for the case of Type IIA
CY orientifold compactifications with background D6-branes wrapping 3-cycles
in the CY. In the simplest configurations one has four SM stacks of
D6-branes labeled ${\bf a},{\bf b}, {\bf c}, {\bf d}$
which correspond  to $U(3)$, $SU(2)$ (or $U(2)$), $U(1)_R$ and $U(1)_L$
gauge interactions respectively, which contain the SM group.
One can construct compactifications with the MSSM particle spectrum in which quarks and leptons lie at the intersections of those SM D6-branes. 
Then the relevant instantons correspond to euclidean D2-branes wrapping 3-cycles in the CY (satisfying specific properties so as to lead to the appropriate superspace interaction). The D2-D6 intersections lie the additional fermionic zero modes which are charged under $(B-L)$. For instantons with the appropriate number of intersections  with the appropriate D6-branes, and with open string disk couplings among the zero modes and the $\nu_R$ chiral multiplet (see fig.(\ref{triangulito})), the operator  
in (\ref{quebonito}) is generated.

The fact that the complex modulus  $U$ transforms under  $U(1)_{B-L}$ gauge 
transformations indicates that  the $U(1)_{B-L}$ gauge boson gets a 
mass from a St\"uckelberg term. So a crucial ingredient in the mechanism to generate non-perturbative masses for the $\nu_R$'s is that there should be  massless $U(1)_{B-L}$ gauge boson which become massive by a St\"uckelberg term. It turns out that not many semi-realistic models with $U(1)_{B-L}$ mass from St\"uckelberg couplings have been constructed up to date. 
In the literature there are two main classes of RR tadpole free models with massive B-L. The first class are non-susy type IIa toroidal orientifold models first constructed in \cite{imr}. The second class are the type II Gepner orientifold models constructed by one of the 
present authors and collaborators \cite{schell1,schell2}. The former were already considered in \cite{iu}. In the present paper we will concentrate on the RCFT Gepner model constructions, which lead to a large class of MSSM like models, more representative of the general Calabi-Yau compactifications
(for a recent discussion of instanton-induced neutrino masses in a model with no RR tadpole cancellation, see \cite{Cvetic:2007ku}).

The class of constructions in \cite{schell1,schell2} start with any of the 168 
Type II compactifications obtained by tensoring $N=2$ SCFT minimal models.  
In addition one can choose a number of modular invariant partition functions (MIPF),
leading to a total of 5403. Then different consistent orientifold projections 
are performed on the different models. This yields a total of 49304 Type II
orientifolds. The open string sector of the theory is defined in terms of
the boundary states of the theory. Intuitively, they play the same role as 
D-branes wrapping cycles in the geometrical settings. Thus one associates 
boundary states   ${\bf a},{\bf b}, {\bf c}, {\bf d}$
to the gauge groups giving rise to the SM. Different choices for the 
SM boundary states lead to different spectra. In the present paper we will 
make use of the data in \cite{schell1} which contains 211634 different MSSM-like 
spectra (including also different hidden sectors).
 Although this number is huge, most of these models are really 
extensions of the MSSM, since they have either   an extra $U(1)_{B-L}$ or 
$SU(2)_R\times U(1)_{B-L}$ group factor beyond the SM group.
As we said, we are actually only interested in models in which 
the $U(1)_{B-L}$ gets a St\"uckelberg mass. Then we find that the number 
of MSSM-like models with these characteristics is dramatically reduced: only 
$0.18$ percent of the models  (391) have a massive $U(1)_{B-L}$. 

As we said, in the geometrical setting of IIA orientifolds with intersecting D6-branes \cite{bgkl,afiru} (see \cite{interev} for reviews and \cite{bachas,aads} for the IIB counterparts),
instantons are associated to D2-branes wrapping 3-cycles, like the background D6-branes do.
Analogously, in the RCFT setting the
same class of boundary states appearing in the SM constructions are the ones
corresponding to  instantons. The zero modes on the instanton is computable from the 
overlaps of instanton brane boundary states (zero modes uncharged under the 4d gauge group)
or of instanton and 4d spacefilling brane boundary states (zero modes charged under the 
corresponding gauge factor).
We find that the criteria for a non-perturbative superpotential to
be generated \cite{Witten:1996bn} are only fulfilled
if the   Chan-Paton (CP) symmetry of the instantons is  $O(1)$. 
For   instantons with CP symmetry \footnote{We adopt the convention that the fundamental representation of $Sp(m)$ is $m$-dimensional.}
$Sp(2)$ or $U(1)$ we find that there are a few extra uncharged fermionic zero modes 
which would preclude the formation of the searched superpotentials.
On the other hand we argue that 
 the addition of fluxes and/or possible non-perturbative extensions
of the orientifold compactifications would allow also instantons with $Sp(2)$ and $U(1)$
symmetries  to generate such superpotentials. We thus include all $O(1)$, $Sp(2)$ and
$U(1)$ instantons 
\footnote{We refer to the different kinds of instanton by their Chan-Paton symmetry on their volume. Since we are not interested in gauge theory instantons, this notation should not be confusing.}
in our systematic search.
The computation of charged and uncharged fermion zero modes 
may be easily implemented as a routine in a systematic computer search for 
instanton zero modes in Gepner MSSM-like orientifolds.
 Results of such a systematic computer search  
are presented in this article.

We find that out of the 391 models with massive $U(1)_{B-L}$,
there are very few  admitting instantons with the required minimal $O(1)$ CP symmetry, and in fact none of them 
without additional vector-like zero modes.
On the other hand we do find 32  models admitting $Sp(2)$ symmetric instantons with just the required charged zero mode content (and the minimal set of non-chiral fermion zero modes).  In fact they
are all variations of the same orientifold  Gepner model based
on the tensor product $(k_1,k_2,k_3,k_4)=(2,4,22,22)$. 
These models all in fact correspond to the same MIPF and orientifold projection,
they only differ on which particular boundary states corresponding to the four  
${\bf a},{\bf b}, {\bf c}, {\bf d}$ SM `stacks'. All models have the same chiral
content, exactly that of the MSSM , with extra vectorlike chiral fields which
may in principle become massive in different points of the CY moduli space.
They have no hidden sector, i.e., the gauge group is just that of the SM.
For each of the models there are 8 instantons with $Sp(2)$ CP symmetry with just the
correct charged zero mode structure allowing for the superpotential coupling 
(\ref{quebonito}) giving rise to $\nu_R$ Majorana masses.
As we said, they have extra  uncharged fermion zero modes beyond the two required 
to generate a superpotentials. However  one would expect that 
these unwanted zero modes might be lifted in more generic situations in which
e.g. NS/RR fluxes are added. 

We thus see that, starting with a 'large' landscape of 211634 MSSM-like models,
and searching for instantons inducing neutrino masses, we find 
there are none admitting the $O(1)$ instantons with exactly the required zero mode structure, and only  few (32) examples with $Sp(2)$ instantons with next-to-minimal uncharged zero mode structure (and exactly the correct charged zero modes).
Let us emphasize though that 
it is the existence of  massive $U(1)_{B-L}$ models which is rare. 
Starting with the subset of models with a massive $U(1)_{B-L}$, 
 finding  models with correct  instanton charged zero modes 
 within  that class is
relatively frequent, 10 percent of the cases. Furthermore, 
we will see  that there are further models
beyond those 32 which contain extra  non-chiral instanton zero modes and which
may also be viable if these modes get massive by some effect (like e.g. the presence  of RR/NS fluxes).

Instantons may generate some other interesting superpotential couplings 
in addition to $\nu_R$ masses, some possibly beneficial and others 
potentially dangerous. In particular we find that in the models which
contain $Sp(2)$ instantons which might induce  
$\nu_R$ masses, there are also other instantons which 
would give rise directly to the Weinberg operator \cite{weinberg}
\beq
{\cal L}_W\ =\ \frac {\lambda }{M} (L{\overline H}L{\overline H})
\eeq
Once the Higgs field gets a  vev,  this gives rise 
directly to left-handed neutrino masses. Thus we find that in that class of models 
both the see-saw mechanism (which also gives rise to a contribution to the Weinberg operator)
and an explicit Weinberg operator might contribute to the physical masses of 
neutrinos. Which effect dominates will depend on the relative size of the 
corresponding instanton actions as well as on the size of the string scale. 
Among potentially dangerous operators which might be generated
stand the R-parity violating operators of dimension $<\ 5$,
which might give rise e.g. to fast proton decay. We make an study 
of the possible generation of those, and find that in all models in which $\nu_R$ masses might be  generated R-parity is exactly conserved. This is a very encouraging result.

A natural question to ask is whether one can say something about the structure 
of masses and mixings for neutrinos. As argued in \cite{iu} 
generically large mixing angles are expected, however to be more quantitative 
we also need to know the structure of Yukawa couplings for leptons. In principle those
may be computed in CFT but in practice this type of computation has not yet been developed 
for CFT orientifolds. Nevertheless we show that, in the case of instantons with $Sp(2)$  
CP symmetry,  a certain factorization  of the flavor structure takes place, which 
could naturally give rise to a hierarchical structure of eigenvalues for neutrino masses.

The structure of this  article is as follows. In the next section we 
present a discussion of instanton induced superpotentials in Type II orientifolds.
This discussion will apply both to Type IIA  and Type IIB CY orientifolds as 
well as to more abstract CFT orientifolds. We discuss the structure of both 
uncharged and charged instanton zero modes.  In particular we show that only instantons 
with $O(1)$ CP symmetry  have the appropriate uncharged zero mode
content to induce a superpotential contribution. We also discuss how
$Sp(2)$ and $U(1)$ might still generate superpotential contributions if
extra ingredients are added to the general setting.
In section 3 we apply that discussion to the 
specific case of the generation of $\nu_R$ Majorana masses, showing what is the required 
zero mode structure in this case. We show how the flavor structure of the 
Majorana mass term factorizes in the  case of instantons with $Sp(2)$ CP symmetry,
leading potentially to a hierarchical structure of eigenvalues.
We further discuss the generation of other B/L-violating operators including the
generation of the Weinberg operator as well as R-parity violating couplings.    
In section 4 we review the RCFT Type II orientifold constructions in \cite{schell1,schell2}.
A general discussion of zero fermion modes for instantons in RCFT orientifolds 
is presented in section 5. 

The results of our general search for instantons generating $\nu_R$ masses
are presented in section 6. We provide a list of all Gepner orientifolds which 
admit instanton configurations potentially giving rise to $\nu_R$ Majorana masses.
We describe the structure of the models with $Sp(2)$ instantons 
 having the  required charged zero modes
for that superpotential to be generated. We also describe  the boundary states of the 
corresponding instantons and  provide the  massless spectrum of the relevant MSSM-like
models. 
Other orientifolds with zero mode structure close to the minimal one
are also briefly discussed. A brief discussion about the possible generation of 
R-parity violating superpotentials is included. 
We leave section 7 for some final comments. Some notation on the CFT orientifold 
constructions, and a discussion of the CFT symmetries in the $Sp(2)$ examples are provided  in two appendices. 
 
As this paper was ready for submission, we noticed \cite{Argurio:2007vq,Bianchi:2007wy}, 
whose discussion of instanton zero modes partially overlaps with our analysis in Section \ref{zeromode}.

\section{Instanton induced superpotentials in Type II orientifolds}

In this Section we review the generation of superpotentials involving 4d charged fields via D-brane instantons in type II compactifications. The discussion applies both to type IIA and IIB models, and to geometrical compactification as well as to more abstract internal CFT's. For recent discussions on D-brane instantons, see \cite{bcw,iu,Florea:2006si,Cvetic:2007ku} .

Before starting, a notational remark in in order. Our notation is adapted to working in the covering theory, namely the type II compactification, and orientifolding in a further step. Thus we describe the brane configurations as a system of branes (described by boundary states for abstract CFT's), labeled $k$, and their orientifold images labeled $k'$. Similarly, we denote $M$ the brane / boundary state corresponding to the instanton brane, and $M'$ its orientifold image. If a brane is mapped to itself under the orientifold action, we call it a `real' brane / boundary, and `complex' otherwise.

\subsection{D-brane instantons, gauge invariance and effective operators}

A basic feature of type II orientifold compactifications with D-branes is 
the generic presence of St\"uckelberg couplings between the $U(1)$ gauge 
fields on the D-branes, and certain 4d RR closed string 2-forms. These 
couplings are required by the Green-Schwarz mechanism when the $U(1)$'s have non-zero triangle contributions to mixed anomalies 
\cite{Sagnotti:1992qw,Ibanez:1998qp}, but can also exist for certain non-anomalous $U(1)$'s \cite{imr,Antoniadis:2002cs}.
These couplings make the $U(1)$ gauge bosons massive, but the symmetry remains as a global symmetry exact in perturbation theory. Since the closed string moduli involved are scalars (0-forms) in
the RR sector, the natural candidate non-perturbative effects to violate these $U(1)$ symmetries are instantons arising from euclidean D-branes coupling to these fields. 

In computing the spacetime effective interaction mediated by the instanton, one needs to integrate over the instanton zero modes. In the generic case (and in particular for the case of our interest) there are no bosonic zero modes beyond the universal ones (namely, the four translational bosonic zero modes associated to the position of the instanton). On the other hand, the structure of fermion zero modes will be crucial. Since we are interested in models with non-trivial 4d gauge group, arising from a set of background 4d spacetime filling branes, we consider separately fermion zero modes which are uncharged under the 4d gauge group and those which are charged.
In this paper we restrict our discussion to 4d $N=1$ supersymmetric models, and this will simplify the analysis of zero modes. 

Fermion zero modes which are uncharged under the 4d gauge group determine the kind of 4d superspace interaction which is generated by the instanton. We are interested in generating superpotential interactions, which, as is familiar, requires the instanton to have two fermion zero modes to saturate the $d^2\theta$ superspace integration. For this, a necessary (but not sufficient!) condition is that the D-branes are half-BPS, so these fermion zero modes are the Goldstinos of the two broken supersymmetries. In the string description, uncharged zero modes arise from open strings in the $MM$ sector  (in our notation, the one leading to adjoint representations), which in particular contain these Goldstinos, and the $MM'$ sector (in our notation, the one leading to two-index symmetric or antisymmetric tensors). Note that both are the same for `real' branes. Hence we are primarily interested in D-branes whose $MM$ sector contains just two fermion zero modes, and whose $MM'$ sector (for `complex' branes) does not contain additional fermion zero modes.

In analogy with the familiar case of gauge theory instantons 
\cite{Hooft:1976fv},
charged fermion zero modes determine the violation of perturbative global symmetries by the instanton-induced interaction. Namely, in order to saturate the integration over the charged fermions zero modes, the spacetime interaction must contain insertions of fields charged under the 4d gauge symmetry, and in particular under the global $U(1)$ factors, which are thus violated by the D-brane instanton. Notice that this holds irrespectively of the number of uncharged fermion zero modes, namely of the kind of superspace interaction induced by the instanton.
Restricting to superpotential interactions, the resulting operator in the 4d effective action has roughly the form
\beqa
W_{inst}\; =\; e^{-U} \, \Phi_1\ldots \Phi_n
\label{thevertex}
\eeqa
Here the  fields $\Phi_1,\ldots, \Phi_n$ are 4d $N=1$ chiral multiplets 
charged under the 4d gauge group, and in particular also under the $U(1)$ 
symmetries. Note also that the instanton amplitude contains a prefactor 
(which in general depends on closed and open string moduli) arising from 
the Gaussian path integral over (massive) fluctuations of the instanton 
(hence described by an open string annulus partition function, see 
\cite{Baumann:2006th,Blumenhagen:2006xt} for related work), which we can 
ignore for our purposes in this paper.

For D-brane instantons, $U$ is the closed string modulus to which the 
euclidean D-brane couples. In the D-brane picture, instanton fermion zero 
modes charged under the gauge factor carried by the $k^{th}$ stack of 4d 
space-filling branes (and its image $k'$) arise from open strings in the 
$Mk$ and $Mk'$ sectors, transforming as usual in the 
$(\fund_M,\antifund_k)$ and $(\fund_M,\fund_k)$ representations, 
respectively (with both related in the case of `real' branes). The (net) 
number of instanton fermion zero modes with such charges is given by 
certain multiplicities \footnote{In geometric type IIA compactifications 
with 4d spacetime-filling branes and instanton branes given by D6- and 
D2-branes wrapped on Special Lagrangian 3-cycles, $I_{Mk}$ corresponds to 
the intersection number between the 3-cycles corresponding to the 
$k^{th}$D6- and the D2-brane $M$ (and similarly for $I_{Mk'}$). In 
geometric type IIB compactifications, it corresponds to the index of a 
suitable Dirac operator. In general (even for abstract CFT's) it can be 
defined as the Witten index for the 2d theory on the open string with the 
boundary conditions corresponding to the two relevant branes. We will 
often abuse language and  refer to this quantity as intersection number, 
even in Section  \ref{scan} where we work in the non-geometric regime of 
type IIB compactifications.}
$I_{Mk}$, $I_{Mk'}$. 

A D-brane instanton, irrespectively of the superspace structure of the 4d 
interactions it may generate, violates $U(1)_k$ charge conservation by an amount $I_{Mk}-I_{Mk'}$ for `complex' branes and $I_{Mk}$ for `real' branes. In particular, this is the total charge of the field theory operator $\Phi_1\ldots\Phi_n$ in (\ref{thevertex}).
From the St\"uckelberg couplings, it is possible to derive 
\cite{iu} (see \cite{Blumenhagen:2006xt,Haack:2006cy,Florea:2006si} for related work, also \cite{Bianchi:2007fx})

 that  
for `complex' instantons, gauge transformations of the $U(1)_k$ vector 
multiplets $V_k\to V_k+\Lambda_k$, transform $U$ as
\beqa
U\to U+ \sum_k N_k (I_{Mk}-I_{Mk'}) \Lambda_k
\label{complexshift}
\eeqa
For `real' brane instantons, which were not considered in \cite{iu}, the shift is given by
\footnote{Equivalently, one may use (\ref{complexshift}), but must include an additional factor of $1/2$ from the reduction of the volume for a real brane (which is invariant under the orientifold action).} 
\beqa
U\to U + \sum_k N_k I_{Mk} \Lambda_k
\label{realshift}
\eeqa
(this new possibility will be an important point in our instanton scan in Section \ref{scan}).

The complete interaction (\ref{thevertex}) is invariant under the $U(1)$ gauge symmetries. However, from the viewpoint of the 4d low-energy effective field theory viewpoint, it leads to non-perturbative violations of the perturbative $U(1)$ global symmetries, by the amount mentioned above.

In the string theory construction there is a simple microscopic explanation for the appearance of the insertions of the 4d charged fields (related to the mechanism in \cite{Ganor:1996pe}). The instanton brane action in general contains cubic terms $\alpha \,\Phi\, \gamma$, involving two instanton fermions zero modes $\alpha$ in the $(\fund_M,\antifund_k)$ and $\gamma$ in the $(\fund_p,\antifund_M)$ coupling to the 4d spacetime field $\Phi$ in the $(\fund_k,\antifund_p)$ of the 4d gauge group \footnote{\label{chirality} Although there is no chirality in $0+0$ dimension, the fermion zero modes $\alpha$ and $\gamma$  are distinguished by their chirality with respect to the $SO(4)$ global symmetry of the system (which corresponds to rotations in the 4d spacetime dimensions, longitudinal to the space-filling branes, and transverse to the instanton brane). Supersymmetry of the instantons constrains the couplings on the instanton action (such as the cubic couplings above) to have a holomorphic structure.}. 
Performing the Gaussian path integral over  the instanton fermion zero modes leads to an insertion of $\Phi$ in the effective spacetime interaction. In general, and for a `complex' instanton, there are several fermion zero modes $\alpha_i$, $\gamma_i$ in the fundamental (resp. antifundamental) of the instanton gauge group,  coupling to a 4d spacetime chiral operators ${\cal O}_{ij}$ (which could possibly be elementary charged fields, or composite chiral operators). Gaussian integration over the fermion zero modes leads to an insertion of the form $\det {\cal O}$ (for `real' brane instantons, $\det{\cal O}$ should be interpreted as a Pfaffian). It is straightforward to derive our above statement on the net charge violation from this microscopic mechanism.

Note that the above discussions show that instantons in different topological sectors (namely with different RR charges, and different intersection numbers with the 4d spacefilling branes) contribute to different 4d spacetime operators. In particular, multiwrapped instantons, if they exist as BPS objects, contribute to operators different from the singly wrapped instanton. This implies that the instanton expansion for a fixed operator is very convergent, and could even be finite. 

Another  important implication of the above discussion is that, in order to generate a specific operator via an instanton process, a necessary condition is that the instanton has an appropriate number and structure of charged zero modes. However, this is not sufficient. Insertions of 4d fields appear only if the fields couple to the instanton fermion zero modes via terms at most quadratic in the zero modes. In equivalent terms, only zero modes appearing in the Gaussian part of the instanton action can be saturated by insertions of 4d fields (those to which they couple). The requirement that the zero modes have appropriate couplings to the 4d fields may be non-trivial to verify in certain constructions. This is the case for the Gepner model orientifolds in coming sections, whose couplings are computable in principle, but unknown in practice.
In such cases we will assume that any coupling which is not obviously forbidden by symmetries will be non-vanishing. Unfortunately there are no arguments to estimate the actual values of such non-vanishing couplings, hence we can argue about the existence of certain instanton induced operators, but not about the coefficients of such terms.

\subsection{Zero mode structure for D-brane instantons}
\label{zeromode}

In this section we describe more concretely different kinds of instantons and the structure of interesting and unwanted zero modes. Our discussion will be valid for general D-brane models in perturbative type II orientifolds without closed string fluxes, although we also make some comments on more general F-theory vacua and the effects of fluxes. A more specific discussion is presented in Section \ref{themess}.

\subsubsection{Uncharged zero modes}

We start discussing zero modes uncharged under the 4d gauge group. These are crucial in determining the kind of superspace interaction induced by the instanton on the 4d theory.
In particular, we are interested in instantons contributing to the 4d superpotential, namely those which contain just two fermion zero modes in this sector. We are also interested in instantons which may contain additional fermion zero modes, and the possible mechanisms that can be used to lift them. Let us discuss `real' and `complex' brane instantons in turn.

\smallskip

$\bullet$  {\bf Real brane instantons} 

Real brane instantons correspond to branes which are mapped to themselves by the orientifold action, hence $M= M'$. Uncharged zero modes arise from the $MM$ open string sector.
As discussed in Section \ref{themess}, there is a universal sector of zero modes, in the sense that it is present in any BPS D-brane instanton, which we now describe. Before the orientifold projection, we have a gauge group $U(n)$ on the volume of $n$ coincident instantons. Notice that, although there are no gauge bosons in $0+0$ dimensions, the gauge group is still well-defined, since it acts on charged states (open string ending on the instanton brane). There are four real bosonic zero modes  and four fermion zero modes in the adjoint representation. For the minimal $U(1)$ case, the four bosons are the translational Goldstones. The four fermions arise as follows. This sector is insensitive to the extra 4d spacefilling branes, and feels an accidental 4d $N=2$ supersymmetry. The BPS D-brane instanton breaks half of this, and leads to four Goldstinos, which are the described fermions \footnote{We thank F. Marchesano for discussions on this point.}.

The orientifold projection acts on this universal sector as follows (see Section \ref{themess} for further discussion). The gauge group is projected down to orthogonal or symplectic.  Hence instanton branes with symplectic gauge group must have even multiplicity (a related argument, in terms of the orientifold action on Chan-Paton indices, is given in Section \ref{themess}). For instantons with $O(n)$ gauge symmetry, the orientifold projects the four bosonic zero modes and two fermion zero modes (related by the two supercharges of 4d $N=1$ supersymmetry broken by the instanton) to the two-index symmetric representation, and the other two fermion zero modes (related by the other two supercharges of the accidental 4d $N=2$ in this sector) to the antisymmetric representation. Hence for $O(1)$ instantons (namely instantons with $O(1)$ gauge group on their volume), we have just two fermion zero modes, which are the Goldstinos of 4d $N=1$ supersymmetry, and the instanton can in principle contribute to the superpotential (if no additional zero modes arise from other non-universal sectors). For instantons with $Sp(n)$ gauge symmetry, the orientifold projects the four bosonic zero modes and two fermion zero modes to the two-index antisymmetric representation, and the other two fermion zero modes to the symmetric representation. Hence, even for the minimal case of $Sp(2)$ instantons, we have just two fermion zero modes in the triplet representation, in addition to the two 4d $N=1$ Goldstinos. Hence $Sp(2)$ instantons cannot contribute to the superpotential in the absence of additional effects which lift these zero modes (see later) \footnote{\label{gauge} For D-brane instantons corresponding to gauge instantons, the additional fermion zero modes in the universal sector couple to the boson and fermion zero modes from open strings stretched between the instanton and the 4d spacefilling brane. They act as Lagrange multipliers which impose the fermionic constraints in the ADHM construction \cite{Billo:2002hm}, and may not spoil the generation of a superpotential.}.

In addition to this universal sector, there exist in general additional modes, whose presence and number depends on the detailed structure of the branes. Namely, on the geometry of the brane in the 6d compact space in geometric compactifications, or on the boundary state of the internal CFT in more abstract setups. They lead to a number of boson and fermion zero modes in the symmetric or antisymmetric representation. The computation of these multiplicities in terms of the precise properties of the instanton branes is postponed to Section \ref{themess}. In order to generate a superpotential, one must require these modes to be absent, except for the case of antisymmetrics of $O(1)$ instantons, which are actually trivial. 

\medskip

An important point is that extra fermion zero modes (including the extra triplet fermion zero modes in the universal sector of $Sp(2)$ instantons, and any two-index tensor fermion zero mode in the non-universal sectors) are in principle not protected against acquiring non-zero masses once the model is slightly modified. In other words, such fermions are non-chiral, in terms of the $SO(4)$ chirality in footnote \ref{chirality}. One such modification is motion in the closed string moduli space, which can lift the non-universal modes  if there are non-trivial couplings between them and closed string moduli (unfortunately, such correlators are difficult to compute, even in cases where the CFT is exactly solvable, like the Gepner models). Note that extra zero modes in the universal sector of $Sp(2)$ instantons cannot be lifted by this effect, since it does not break the accidental 4d $N=2$ in this sector. A second possible modification which in general can lift extra zero modes  is the addition of fluxes, generalizing the results for D3-brane instantons in geometric compactifications \cite{fluxinst} (for non-geometric CFT compactifications, we also expect a similar effect, once fluxes are introduced following \cite{becker:2006ks}). Note that fluxes can lift extra zero modes in the universal sector as well, since they can break the accidental 4d $N=2$ susy in this sector. A last mechanism arising in more general F-theory compactifications and discussed below for complex instantons, is valid for real instanton branes as well.

The bottom line is that in the absence of such extra effects, only $O(1)$ instantons can contribute to superpotential terms. However, in modifications of the model such extra effects can easily lift the extra fermion zero modes. Hence, this kind of extra vector-like zero modes will not be considered catastrophic, and real instantons (including the $O(1)$ and $Sp(2)$ cases) with such zero modes are considered in our scan in Section \ref{scan}.

\medskip
\medskip
\medskip
\medskip

$\bullet$ {\bf Complex brane instantons} 

Zero modes uncharged under the 4d gauge group can arise from the $MM$ and $MM'$ open string sectors. Notice that the orientifold action maps the $MM$ sector to the $M'M'$, hence we simply discuss the former and impose no projection. The discussion of the $MM$ sector is as for real brane instantons before the orientifold projection. The universal sector leads to a $U(n)$ gauge symmetry, and four bosonic and four fermionic zero modes in the adjoint representation. The bosons are translational Goldstones, while the fermions are Goldstinos of the accidental 4d $N=2$ present in this sector.
Hence, even in the minimal case of $U(1)$ brane instantons there are two extra fermion zero modes, beyond the two fermion zero modes corresponding to the 4d $N=1$ Goldstinos. Hence $U(1)$ instanton (except for those corresponding to gauge instantons, see footnote \ref{gauge}) cannot contribute to superpotential terms in the absence of additional effects, like closed string fluxes . However, keeping in mind the possibility of additional effects lifting them in modifications of the model, we include them in the discussion. Also, in what follows we will use the $U(n)$ notation for the different fields to keep track of the Chan-Paton index structure.

The above statement would seem in contradiction\footnote{We thank S. Kachru for discussions
 on the ideas in this paragraph.}
 with computations of non-perturbative superpotentials \cite{Witten:1996bn}
 induced by M5-branes instantons in M-theory compactifications on Calabi-Yau fourfolds,
 which are dual to D3-brane instantons (with world-volume $U(1)$ gauge group) on type IIB compactifications.
 The resolution is that the M5-branes that contribute are intersected 
by different $(p,q)$ degenerations of the elliptic fiber.
 This implies that $U(1)$ D3-brane instanton only contribute
 if they are intersected by mutually non-local $(p,q)$ 7-branes.
 The two extra fermion zero modes exist locally on the D3-brane volume,
 but cannot be defined globally due to the 7-brane monodromies.
 Hence such effect can take place only on non-perturbative
 type IIB compactifications including $(p,q)$ 7-branes.
 Note that in perturbative compactifications,
 namely IIB orientifolds, the $(p,q)$ 7-branes are
 hidden inside orientifold planes \cite{sen:1996vd} with their monodromy 
encoding the orientifold projection; hence the only branes that can contribute 
to the superpotential are real branes, for which the projection/monodromy 
removes the extra fermion zero modes, as discussed above.
 
In addition to this universal sector, the $MM$ sector may contain a non-universal set of fermions and bosons, in the adjoint representation (hence uncharged under $U(1)$). They depend on the specific properties of the brane instanton, and will be discussed in Section \ref{themess}. These additional zero modes should be absent in order for the instanton to induce a non-trivial superpotential. Notice however that these zero modes are uncharged under any gauge symmetry, and hence vector-like. Thus, there could be lifted in modifications of the model, as discussed for real instantons.

The $MM'$ sector is mapped to itself under the orientifold action. Hence it leads to a number of bosons and fermions in the two index symmetric or antisymmetric representations. Notice that the two-index antisymmetric representation is trivial for $U(1)$, so these modes are actually not present. On the other hand, fermion zero modes in the two-index symmetric representation are chiral and charged under the brane instanton gauge symmetry. Hence they cannot be lifted by any of the familiar mechanisms, and thus spoil the appearance of a non-perturbative superpotential, even if the model is modified.
Such fermion zero modes are considered catastrophic and we will look for models avoiding them in our scan in Section \ref{scan}.
 
\subsubsection{Charged fermion zero modes} 
 
$\bullet$ {\bf Real brane instantons}

Instanton zero modes charged under the 4d gauge group arise from $Mk$ 
open string sectors (and their image $Mk'$). In the generic case, there 
are no scalar zero modes in these sectors. This is because in mixed $Mk$ 
open string sectors the 4d spacetime part leads to DN boundary 
conditions, which already saturate the vacuum energy in the NS sector. 
Only in the special case where the internal structure of the 
spacetime filling brane $k$ and the instanton brane are the same, there 
may be NS ground states of the internal CFT which do not contribute extra 
vacuum energy, hence leading to massless scalars. However, this 
corresponds to brane instantons which can be interpreted as instantons of 
the 4d gauge theory on the 4d space-filling branes. The instantons we are 
interested in for the generation of neutrino Majorana mass terms are not 
of this kind \cite{iu} (see e.g.
\cite{Bershadsky:1996gx,Acharya:2000gb,Billo:2002hm,Florea:2006si,Akerblom:2006hx}
for discussions on gauge theory instantons from D-brane instantons).

Hence we focus on charged fermion zero modes, which are generically present in any mixed $Mk$ sector. Let us define $L_{Mk}$, $L_{Mk'}$ the (positive by definition) number of left-handed chiral fermion zero modes in the representations $(\fund_M,\fund_k)$, $(\fund_M,\antifund_k)$, respectively. The net number of chiral fermion zero modes in the $(\fund_M,\antifund_k)$ is given by $I_{Mk}=L_{Mk'}-L_{Mk}$. This controls the violation of the $U(1)_a$ global charge by the instanton. Namely, such fermion zero modes in the $Mk$, $Mp$ sectors lead (if suitable couplings are present) to the insertion of 4d charged fields $\Phi_{kp}$ and/or $\Phi_{kp'}$. 

In addition, there are $P_{Mk}={\rm min}(L_{Mk'}, L_{Mk})$ vector-like pairs of fermion zero modes. Since they are vector-like, they may be lifted by slight modifications
 of the model, like moving in the closed string moduli space, or by introducing additional ingredients, like fluxes.
 In addition, they may be lifted by moving in the open string moduli space of the 4d spacefilling branes, as follows.
 The zero modes may lead to insertions of 4d fields $\Phi_{kk}$, if the $kk$ sector contains such 4d chiral multiplets (or to insertions of composite 4d operators in the adjoint of the $k^{th}$ 4d gauge factor), and if they couple to the zero modes. Although this may not be generically not the case, many of our models in coming section contain such adjoint fields. Hence, a non-trivial vev for the latter can lift these extra vector-like zero modes, hence leading to instanton generating the superpotential of interest. Given these diverse mechanisms to lift these zero modes, their presence of such zero modes is thus  unwanted, but again not necessarily catastrophic.

\medskip

One last comment, related to the concrete kind of instanton search we will be interested in.
 Namely, we will be searching for instantons leading to a specific operator, carrying non-trivial 
charges under a specific set of 4d gauge factors. Postponing the detailed discussion to 
Sections \ref{mssm}, \ref{gepnersm} , let us denote ${\bf a}$, ${\bf b}$, ${\bf c}$, ${\bf d}$ the set of branes leading to a field theory sector, denoted `observable' (and which reproduces the SM in our examples). We will require the instanton to have a prescribed number of chiral fermion zero modes charged under these branes, namely we require the intersection numbers of the instanton with these branes $I_{M{\bf a}}, \ldots, I_{M{\bf d}}$ to have specific values (as mentioned above, in the most restrictive scan we forbid vector-like pairs of zero modes under these branes). In addition, the model in general contains an additional sector of branes, denoted `hidden' (since there is zero net number of chiral multiplets charged under both sectors) and labeled $h_i$, required to  fulfill the RR tadpole cancellation conditions. 
In general there may be instanton fermion zero modes from e.g. the $Mh_1$, $h_2M$ sectors,
 which would contribute to insertions of the 4d fields in the $h_1h_2$ sector if there are
 suitable cubic couplings. These  extra insertions could be avoided if such
 4d fields in the hidden sector acquire vevs (note that vevs for the (vector-like)
 fields charged under the visible and hidden sectors would typically break hypercharge,
 and should be avoided), and 
hence lift 
the zero modes. Equivalently, from the 4d perspective,  the unwanted extra $h_1h_2$ field insertion is replaced by its vev. 
However, this renders the discussion very model dependent. Moreover, 
the possibility of hidden brane recombination was not included 
in the search for SM-like models in \cite{schell1,schell2} (namely,
 the possibility of allowing for chiral fields charged under the observable 
and hidden gauge groups, which may become non-chiral and possibly massive upon hidden brane recombination).
 Hence we will consider these chiral fermion zero modes as unwanted 
 (as usual, non-chiral modes are unwanted but not catastrophic, hence they are allowed for in 
a more  relaxed search).

$\bullet$ {\bf Complex brane instantons}

The discussion of `complex' brane instantons is somewhat analogous to the previous one, with the only complication that the brane $M$ and its image $M'$ lead to independent modes, leading to a more involved pattern of fermion zero modes. Instanton zero modes charged under the 4d gauge group arise from the $Mk$,$Mk'$ and related sectors. As for `real' brane instantons, there are generically no scalars in these sectors (and certainly not in our case of interest). Hence we focus on charged fermion zero modes, which are generically present in any mixed sector. 

In contrast with `real' brane instantons, a net combination of fermion zero modes in the 
$(\fund_M,\antifund_k)+(\fund_M,\fund_k)$ is not vector-like, 
but chiral under the instanton gauge symmetry. Such a pair cannot therefore be lifted
 even by modifications of the theory. In general, if the instanton has a mismatch in the 
total numbers $n_\alpha$, $n_\gamma$ of fermion zero modes $\alpha_i$ in
 the $\fund_M$ and $\gamma_j$ in the $\antifund_M$, the instanton amplitude
 automatically vanishes. Namely, the matrix of operators ${\cal O}_{ij}$ coupling
 to the zero modes necessarily has rank at most ${\rm min}(n_\alpha,n_\gamma)$.
 That is , if $n_\alpha>n_\gamma$ there are linear combinations of the $\alpha_i$
 which do not couple, and cannot lead to insertions. Moreover, they are not liftable by 
the familiar mechanisms \footnote{Note that such a mismatch is always correlated with
 the existence of extra chiral zero modes in the $MM'$ sectors, discussed above.
 Denoting ${\vec Q}_a$, ${\vec Q}_{\rm orient}$ the vector of RR charges of 
the 4d space-filling branes and orientifold planes, they satisfy the RR tadpole
 conditions $\sum_a N_a {\vec Q}_a +\sum_{a'} N_a {\vec Q}_{a'}+{\vec Q}_{\rm orient.}=0$. 
By taking the `intersection' bilinear with the RR charges ${\vec Q}_{M}$ of the brane instanton,
 we have $I_{Ma}+I_{Ma'}+I_{M,{\rm orient}}=0$. This implies that the number of fundamentals
 minus anti-fundamentals of the instanton gauge group is related to  the number of two-index tensors.},
 thus in our instanton search in Section \ref{scan} such excess zero modes are forbidden even in relaxed scans.

Let us thus discuss a sector of fermion zero modes with equal number $n_\alpha=n_\gamma$. Considering a given 4d space-filling brane $k$, let us denote $L_{Mk}$, $L_{M'k'}$, $L_{Mk'}$, $L_{M'k}$ the (positive by definition) number of left-handed chiral fermion zero modes in the representations $(\fund_M,\fund_k)$, $(\antifund_M,\antifund_k)$, $(\fund_M,\antifund_k)$, and $(\antifund_M,\fund_k)$ respectively. The net number of chiral fermion zero modes in the $(\fund_M,\antifund_k)$ and $(\fund_M, \fund_k)$ is given by $I_{Mk}=L_{Mk'}-L_{M'k'}$ and $I_{Mk'}=L_{Mk}-L_{M'k}$, respectively. This net number of fermions zero modes controls the violation of the $U(1)_a$ global charge by the instanton. Namely, such fermion zero modes in the $Mk$, $Mp$, $Mk'$, $Mp'$ sectors lead (if suitable couplings are present) to the insertion of 4d charged fields $\Phi_{kp}$ and/or $\Phi_{kp'}$. 

The remaining fields in this sector are vector-like pairs, in the $(\fund_M,\antifund_k)+(\antifund_M,\fund_k)$ or the $(\fund_M,\fund_k)+(\antifund_M,\antifund_k)$, which in principle lead to the vanishing of the instanton amplitude, but which can be lifted by additional effects (motion in closed or open string moduli space, or addition of fluxes), in a way consistent with the gauge symmetries in 4d spacetime and on the instanton.

\medskip

Just like for `real' brane instantons, we conclude by commenting on our concrete instanton search in models with a set of visible branes ${\bf a}$, ${\bf b}$, ${\bf c}$, ${\bf d}$ and a set of hidden branes $h_i$. The requirement that the instanton leads to an operator with specific charges under the visible branes  fixes the values of the quantities $I_{M{\bf a}}-I_{M{\bf a'}}$, etc.
As we described for real branes, one may still have fermion zero modes charged under the hidden sector branes, but they lead to additional insertions, hence we rather focus on instantons with $I_{Mh_i}-I_{Mh_j'}=0$. The two kinds of conditions, on intersection numbers with visible and hidden branes, still leave the possibility of combinations of fermion zero modes of the kind
$(\fund_M,\antifund_k)+(\antifund_M,\fund_k)$, which do not contribute to $I_{Mk}$, or of the kind $(\fund_M,\fund_k)+(\antifund_M,\antifund_k)$, which does not contribute to $I_{Mk'}$. Such combinations are automatically vector-like, and thus may be lifted in modifications of the theory.  But the condition also allow combinations like $(\fund_M,\antifund_k)+(\fund_M,\fund_k)$, which exploit a cancellation between $I_{Mk}$ and $I_{Mk'}$ (as also does $(\antifund_M,\fund_k)+(\antifund_M,\antifund_k)$). Such combinations are chiral by themselves, and in general imply a mismatch of modes in the $\fund_M$ and the $\antifund_M$. The total mismatch can be arranged to vanish using combinations of the kind 
$(\fund_M,\antifund_k)+(\fund_M,\fund_k)$ and $(\antifund_M,\fund_p)+(\antifund_M,\antifund_p)$ for different branes. 
However, the only way to lift these pairs is by breaking the gauge symmetry on the 4d space-filling branes $k$ and $p$.
This can be done without damage to the visible sector if these are hidden branes, but this corresponds to the recombination of hidden branes that, as mentioned already, we are not going to consider. Hence only vector-like pairs with respect to {\em each} brane are considered to be liftable in simple modifications of the theory.
In our instanton search, these are the only additional fermion zero modes which we allow in relaxed scans (but they are clearly not allowed for in restricted scans)

\section{Instanton induced Majorana neutrino masses}
\label{majoranas}

In this Section we discuss the possible physical effects of D-brane instantons in
 string models with SM-like spectrum. In particular we describe the conditions
 to generate right-handed neutrino Majorana masses. We also comment on other possible
 $B$ and/or  $L$ violating operators that can be generated by instantons.
In this section we will again use the geometrical language of IIA intersecting 
D-branes but it should be clear that our discussion equally applies to general 
CFT orientifolds like the ones presented in the next section.

\subsection{The MSSM on the branes}
\label{mssm}

\begin{figure}
\centering
\epsfxsize=3.5in
\hspace*{0in}\vspace*{0in}
\epsffile{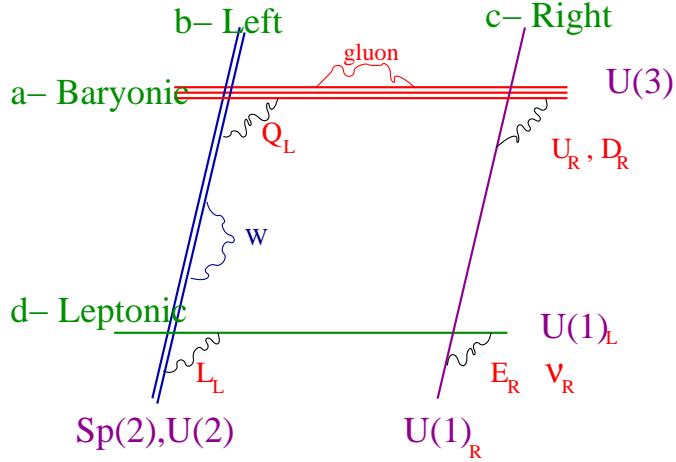}
\label{madrid}
\caption{\small Quarks and leptons at intersecting branes }

\end{figure}

Let us now specify the discussion in 
the previous section to the case of the generation of a right-handed neutrino 
mass term. In order to do that we
need some realistic orientifold  construction
with the gauge group and fermion spectrum of the 
Standard Model (SM). 
In the context of Type II orientifolds perhaps the most economical 
brane configuration leading to a SM spectrum is the
one first considered in \cite{imr}. This consists of four stacks, labelled ${\bf a,b,c,d}$.  The gauge factor on branes {\bf a}  is $U(3)$, and contains QCD and baryon number. The {\bf d} factor is $U(1)_{\bf d}$, and corresponds to lepton number. Stack {\bf b} contains $SU(2)_{\rm Weak}$ either embedded in $U(2)$ or $Sp(2)$. Finally brane {\bf c} can either provide a $U(1)$ or an $O(2)$ factor.
In the brane intersection language, the chiral fermions of the SM 
live at the intersections of these branes, as depicted in Fig.~\ref{madrid}.

The $U(1)_Y$ factor of the standard model is embedded in the Chan-Paton factors of branes {\bf a,c} and {\bf d} 
as \beq
\label{Ydef}
 {Y}\ = \ \frac 16  {Q_{\bf a}} -\frac 12 Q_{\bf c} - \frac 12 Q_{\bf d} \, =\,
\frac {1}{2}( {Q_{B-L}}- {Q_R})
\eeq
where $Q_{\bf x}$ denotes the generator of the $U(1)$ of brane stack ${\bf x}$ (in case the Chan-Paton factor of brane
{\bf c} is $O(2)$ one should use the properly normalized $O(2)$ generator). Note that in this convention the $Q_d$ generator appears with sign opposite to other conventions in the literature, e.g. in \cite{iu}.
In addition to $Y$ these models have two  additional $U(1)$ gauge symmetries:
\beqa
 {Q_{\rm anom}}\ &  =\ & 3Q_{\bf a}+Q_{\bf d}\, =\, 9 {Q_B}+ {Q_L} \nonumber \\
 {Y'}\ & = & \frac 13 Q_{\bf a} + Q_{\bf c} - Q_{\bf d} \, =\,  {Q_{B-L}} +  {Q_R}
\eeqa
The first is anomalous whereas the second,  which we will call $B-L$ (with a slight abuse of language,
since it is in fact a linear combination of $B-L$ and hypercharge), is anomaly free.
In models in which the electroweak gauge group is 
embedded in $U(2)$, rather than in $Sp(2)$, there is 
a second  anomalous $U(1)_{\bf b}$.
The charges of the SM particles under these $U(1)$ symmetries are given in table~\ref{tablasm}.
\begin{table}[htb] \footnotesize
\renewcommand{\arraystretch}{0.75}
\begin{center}
\begin{tabular}{|c|c|c|c|c|c|c|c|}
\hline Intersection &
 $D=4$ fields/ {zero modes}  &   &  $Q_{\bf a}$  &   $Q_{\bf c} $ & $Q_{\bf d}$  & Y & $Q_M$
\\
\hline\hline ({\bf ab}),({\bf ab'}) & $Q_L$ &  $3(3,2)$ & 1  & 0 & 0 & 1/6   & 0\\
\hline ({\bf ca}) & $U_R$   &  $3( {\bar 3},1)$ &  -1   & 1  & 0 & -2/3  & 0
\\
\hline ({\bf c'a}) & $D_R$   &  $3( {\bar 3},1)$ &  -1    & -1  & 0 & 1/3   & 0
\\
\hline ({\bf db}),({\bf db'}) & $ L$    &  $3(1,2)$ &  0    & 0  & 1 & -1/2  & 0\\
\hline ({\bf c'd}) & $E_R$   &  $3(1,1)$ &  0   & -1  & -1  &  1   & 0 \\
\hline ({\bf cd}) &  {$\nu_R$}   &  $3(1,1)$ &  0   & 1  & -1  & 0  & 0 \\
\hline \hline
 (M{\bf c}) &   {$\alpha_i$}   &  $2(0,0)$ &  0     & -1  & 0 & 1/2  & 1 \\
\hline ({\bf d}M) &  {$\gamma_i$}   &  $2(0,0)$ &   0  & 0  & 1  & -1/2   &  -1 \\
\hline
\end{tabular}
\end{center}
\caption{\small Standard model spectrum and $U(1)$ charges of
particles and zero modes. $Q_M$ stands for the world-volume
gauge symmetry in the case of $U(1)$ complex instantons.}
\label{tablasm}
\end{table}

The $U(1)_k$ gauge symmetries have couplings with the RR 2-forms $B_r$ of the model, as follows
\beq
 {S_{BF}}\, =\, \sum_{k,r} \,
 {N_k}( {p_{kr}}-p_{k'r}) \int_{4d}\,  {B_r} \wedge   {F_k}
\eeq
where $p_{kr}$, $p_{k',r}$ are given by the RR charges of the D-branes.
These imply that {under a $U(1)_k$  gauge transformation 
${ A_k } \to  {A_k } \, + \, {d\Lambda_k}$ the  scalar $a_r$ dual to $B_r$ transforms as
\beq
 {a_r }\to  {a_r } \, +\, \sum_k\,  {N_k}\,  ({p_{kr}}-p_{k'r})\,  {\Lambda_k}
\eeq
This has  { two effects}:
1) The linear combination of axion fields  $ \sum_{r} ( {p_{kr}}-p_{k',r}) {a_r }$
is eaten up by the  $U(1)_k$ massless gauge boson, making it massive.
2) For anomalous $U(1)_k$, the anomalies cancel through a 4d version of the Green-Schwarz mechanism. This works due  to the existence of  appropriate 
$a_r\, F\wedge F$ couplings, involving the  gauge fields in the theory.

It is obvious that all anomalous $U(1)$'s become massive by this mechanism.
However it is important to realize \cite{imr} that  gauge bosons of
  anomaly-free symmetries like $U(1)_{B-L}$ may also become massive by combining with a 
linear combination of axions. This is interesting since it provides a mechanism to
 reduce the gauge symmetry of the model without needing explicit extra Higgsing. 
In the models in which $U(1)_{B-L}$ becomes massive in this way, the gauge group left over is purely that of the SM.  Moreover, we will see that having (B-L) massive by this St\"uckelberg mechanism is crucial to allow the generation of instanton-induced Majorana neutrino masses.

Note that the $B\wedge F$  couplings may also be potentially dangerous, since in principle they could also exist for hypercharge, removing $U(1)_Y$ from the low-energy spectrum. As we will see in our RCFT examples later on, having massless $U(1)_Y$ but massive $U(1)_{B-L}$ turns out to be a strong constraint in model building.

\subsection{Majorana mass term generation}

As discussed in the previous section, string instantons can give rise to
non-perturbative superpotentials breaking explicitly  the perturbative global $U(1)$ symmetries left-over from $U(1)$ gauge bosons made  massive through the St\"uckelberg mechanism. The kind of operator we are interested in has  the form
\beq
W\simeq e^{- {S_{ins}}}\,  {\nu_R\nu_R }
\eeq
where $\nu_R$ is the right-handed neutrino superfield
\footnote{ Actually we denote by $\nu_R$ the left-handed $\nu_L^c$ field following 
the usual (a bit confusing) convention.}.
 Here $S_{ins}$ transforms under both $U(1)_{B-L}$ and $U(1)_R$} in such a way that 
the overall operator is 
gauge invariant. This operator may be created if the mixed open string sectors
 lead to { fermionic zero modes} {$ \alpha_i , \gamma_i $ }, ${i=1,2}$, appropriately 
charged under the 4d gauge factors. As we discussed in the previous section,
to generate a superpotential one needs instanton with $O(1)$ Chan-Paton symmetry,  in order to lead to two uncharged  fermion zero modes to saturate the $d^2\theta$ 4d superspace integration.
On the other hand, as we argued, instantons with $Sp(2)$ or $U(1)$ CP symmetries 
may also induce the required superpotentials if there is some additional dynamics 
getting rid of the extra uncharged zero modes which in principle appear in 
instantons with these symmetries. We thus consider all $O(1)$, $Sp(2)$ and $U(1)$ 
instantons in our discussion.

\begin{figure}
\epsfysize=4.0cm
\begin{center}
\leavevmode
\epsffile{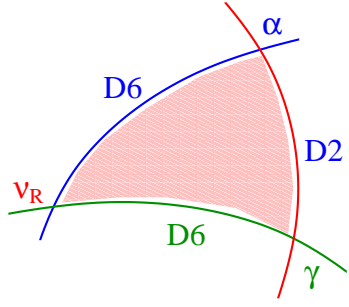}
\end{center}
\label{triangulito}
\caption{\small  Disk amplitude coupling two charged zero modes to $\nu_R$ in the geometrical Type IIA intersecting brane approach. }
\end{figure}

In order to  to get a $\nu_R$ bilinear, the intersection numbers of instanton $M$ and $d,c$ branes are as follows 
\beq
Sp(2)\; {\rm case} :\quad \quad \quad\quad\quad \quad\quad \quad\quad\quad\quad
\label{InstInterOne}
 { I_{M{\bf c}} =1} \ ;\  {I_{M{\bf d}}=-1 } \quad\quad
\eeq
(since there is an extra multiplicity from the two branes required to produce $Sp(2)$)
\beq
O(1)\; {\rm case} :\quad\quad \quad\quad\quad \quad\quad\quad\quad\quad\quad\quad
\label{InstInterTwo}
 { I_{M{\bf c}} =2} \ ;\  {I_{M{\bf d}}=-2 } \quad\quad
\eeq
\beq
U(1)\; {\rm case} :\quad\quad \quad
\label{InstInterThree}
  { I_{M{\bf c}} =2} \ ;\  {I_{M{\bf d}}= -2 }
\ \ {\rm or } \ \ 
\  {I_{M{\bf d'}} = 2} \ ;\  {I_{M{\bf c}'}= -2 }
\eeq
Furthermore  there must be 
cubic couplings involving the right-handed neutrino
superfield  $\nu^a$ in the $a^{th}$ family and the fermionic zero modes $\alpha_i,\gamma_j$
\beq
  {L_{cubic}}\ \propto \ {d^{ij}_a}\ ( {\alpha_i}
  {\ \nu }  {^a \gamma_j}) \ \ , a=1,2,3
\eeq
In type IIA geometric compactifications, this coupling arises from open string disk instantons, see Fig.~\ref{triangulito}. In general type IIA models (resp. IIB models), the coefficients $d_a^{ij}$  depend on the K\"ahler (resp. complex structure) moduli, and possibly on  open string moduli. In simple CFT models (like e.g. in toroidal cases) these quantities may be in principle explicitly computed.
%

These trilinear couplings appear in the instanton action and after integration
of the fermionic zero modes $\alpha_i , \gamma_i$ one gets a 
superpotential coupling proportional to
\beq
\int d^2\theta 
\int  {d^2\alpha\,  d^2\gamma }  \ e^{- {d^{ij}_a}\ (
{\alpha _i } {  \nu^a }
 {\gamma_j})}
\ = \  \int d^2\theta\, {\nu_a \nu_b}\,  (\, \epsilon_{ij}\epsilon_{kl}
 {d^{ik}_ad^{jl}_b}\, )
\quad
\eeq
yielding a  {right-handed neutrino mass term}. This term is multiplied by the exponential of the instanton euclidean action so that the final result for the right-handed neutrino mass  (up to a 1-loop prefactor) has the form
\beq
M^R_{ab}\ =\ 
 {M_s}
 (\,\epsilon _{ij} \epsilon_{kl}
  {d^{ik}_ad^{jl}_b}\, ) \exp(\,-\frac{V_{\Pi_M}}{g_s} +i \sum_r {q_{M,r}} { a_r}\, )
\eeq
For geometric compactifications $V_{\Pi_M}$ is roughly related to the wrapped volume. We keep the same notation to emphasize that the effect is non-perturbative in $g_s$. In supersymmetric models the term in the exponential is the linear combination $U$ of complex structure moduli to which the instanton D-brane couples, as described in the previous section. As explained, the  { gauge $U(1)_{\bf c}$, $U(1)_{\bf d}$ transformation of the bilinear piece and the } $e^{- {S_{D2}}}$  {factor}
 { nicely cancel}. Note that from the viewpoint of the 4d SM effective
field theory, the instanton has generated a Majorana neutrino mass violating $B-L$. Notice 
also that since this symmetry is non-anomalous, its violation cannot be associated to a gauge instanton, hence this 
is a pure string theory instanton effect.

\subsection{Flavor and the special case of $Sp(2)$ instantons}
\label{flavour}

In order to extract more specific results for the flavor structure
of the obtained Majorana mass operator, one needs to know more details
about the quantities $d_a^{ij}$ coming  from the disk correlators.
However in the particular case of $Sp(2)$ instantons, the labels $i,j$ are 
$Sp(2)$ doublet indices, and the symmetry requires $d_a^{ij}=d_a \epsilon^{ij}$. The mass matrix for the
 three neutrinos is given by $M^R_{ab}=$$  2M_s d_ad_b \exp{(-U)}$,
so that the flavour dependence on $a,b=1,2,3$ factorizes. More generally, as we will
 see in our RCFT search in Section \ref{scan}, there are typically several
different instantons contributing to the amplitude, so that we actually have
a result for the mass
\beq
\label{InstSum}
M_{ab}^R\ =\ 2M_s \sum_r \ d_a^{(r)}d_b^{(r)} \  e^{-U_r}
\eeq
where the sum goes over the different contributing instantons. One thus has a structure of the form 
\beq
M^R \ \sim \ \sum_r  \ 
e^{-U_r}
\diag
(d_1^{(r)},d_2^{(r)},d_3^{(r)}) \cdot
\left(
\begin{array}{ccc}
1 & 1 & 1 \\
1 & 1 & 1 \\
1 & 1 & 1
\end{array}
\right)
\cdot 
\diag(d_1^{(r)},d_2^{(r)},d_3^{(r)})  \ \ .
\label{matrilla}
\eeq
This structure is very interesting. Indeed, each instanton makes one particular
 (instanton-dependent) linear combination of the neutrinos massive, 
leaving two linear combinations massless. Hence, for three or more instantons, 
one generically has a matrix with three non-zero eigenvalues. 
It is easy to imagine a hierarchical structure  
among the three eigenvalues if e.g. the exponential suppression 
factors $\exp(-{\rm Re\,} U_r)$ are different for each instanton.

\subsection{Other $B-$ and $L-$violating operators}
\label{otherbl}

Our main focus in this paper is on the generation of right-handed 
neutrino Majorana masses. However instantons may induce other L- and
B-violating operators which we briefly summarize in this subsection.


\subsubsection{\bf The Weinberg operator }

A right-handed neutrino Majorana mass term is not the only
possible operator violating lepton number. Instanton effects
may also give rise to
dimension 5 operators not involving $\nu_R$. Specifically, the Weinberg operator
\beq
{\cal L}_W\ =\ \frac {\lambda }{M} (L{\overline H}L{\overline H})
\label{weinberg} \ \ .
\eeq
might be generated.
Once  Higgs fields get a vev $v$ this operator gives  rise directly to
Majorana masses for the left-handed neutrinos of order $\simeq v^2/M$.
Indeed, it is easy to check that in this case the required
instanton $M$ must verify
\beq
Sp(2) \; {\rm case} : \quad\quad\quad\quad\quad\quad\quad\quad
  { I_{M{\bf c}} = -1} \ ;\  {I_{M{\bf d}}= 1 } \quad\quad\quad\quad\quad\quad
\eeq
\beq
O(1) \; {\rm case} : \quad\quad\quad\quad\quad\quad\quad\quad
  { I_{M{\bf c}} = -2} \ ;\  {I_{M{\bf d}}= 2 } \quad\quad\quad\quad\quad\quad
\eeq
\beq
U(1) \; {\rm case} : \quad\quad   { I_{M{\bf c}} = -2} \ ;\  {I_{M{\bf d}}=2 }
\ \ {\rm or}  \ \
\  {I_{M{\bf c}'} =2} \ ;\  {I_{M{\bf d}'}= -2 }
\eeq
(here we are assuming $SU(2)_{weak}$ to be embedded in an $Sp(2)$).
Note that these intersection numbers are different to those giving rise
to $\nu_R$ mass terms. In particular they lead to a transformation
under $B-L$ opposite to that of $\nu_R$ mass operators
\footnote{Instantons with these intersection numbers will be denoted 
with a plus sign in the instanton search later on}.
In the present case there are altogether four  fermionic zero modes
 $\alpha_i$,$\gamma_i$ corresponding
to the intersections of the instanton M with the
branes ${\bf c}$, ${\bf d}$. These zero modes can have couplings 
involving the  left-handed leptons $L$ and the u-type Higgs
multiplet ${\overline H}$
\beq
L_{disk}\ \propto \ c^{ij}_a \ (\alpha_i  (L^a {\overline H}) \gamma_j)  \ .
\eeq
Upon integration over the fermionic zero modes one recovers
the Weinberg operator. In the present case the scale $M$ of the
Weinberg operator will be the string scale $M_{s}$ and the coupling
$\lambda \simeq \exp(-S_{ins})$. 
Again, in the particular case of $Sp(2)$ instantons 
the situation simplifies ($c^{ij}_a=c^a \epsilon ^{ij}$) 
 and one gets 
 left-handed  neutrino Majorana masses 
\beq
M_{ab}^L\  =  \   \frac { <{\overline H}>^2}
{M_s} 
\sum_r \ 2c_a^{(r)}c_b^{(r)}  \  e^{-S_r}
\eeq
where $r$ runs over the different contributing instantons and
$S_r$ is their corresponding action. The flavour structure 
of this left-handed neutrino mass matrix is the same as in eq.(\ref{matrilla})
and again may potentially lead to a hierarchical structure of 
left-handed neutrino masses, as is experimentally observed.

In a given  model both this kind of instanton and the one
giving rise to right-handed neutrino masses (which is different)
may be present. This contribution to
the left-handed neutrino Majorana mass
is in principle sub-leading compared to the see-saw
 contribution
\beq
M_{ab}^L{\hbox{(see-saw}})\  =  \   \frac { <{\overline H}>^2}
{2M_s}
h_D^T (\sum_r \ d_a^{(r)}d_b^{(r)}  \  e^{-S_r})^{-1}\ h_D
\eeq
where is the ordinary Yukawa coupling constant $h_D^{ab}(\nu_R^a{\bar H}L^b)$.
In principle the former  is doubly suppressed both by
$1/M_s$ and the exponential factor. On the other hand if the
exponential suppression is not too large 
 this mechanism involving directly
the Weinberg operator may be the most relevant source
of neutrino masses. This is because  the see-saw contribution 
coming from $\nu_R$ exchange is proportional to the 
square of the ordinary Yukawa couplings  $h_D^{ab}$
which could be small.
One could even think of having just the Weinberg operator as the 
unique source of the observed left-handed neutrino masses.
 Note however that in string vacua like this, in which
the $\nu_R$'s are present and massless at the perturbative level, 
having just the Weinberg operator would not be phenomenologically correct, and
instantons of the first class are still needed so that the 
$\nu_R$'s get a sufficiently large mass.


\subsubsection{\bf R-parity violating operators }

In the case of $N=1$ SUSY models like the MSSM there might be 
operators of dimension 3 and 4 violating lepton and/or baryon number.
These are the superpotential couplings
\beq
W_{R_p}\ =\ \mu^{L}_a L^a{\overline H} \ +\ \lambda _{abc} Q^aD^bL^c
\ +\ \lambda '_{abc}U^aD^bD^c \ +\ \lambda ''_{abc}L^aL^bE^c
\label{rparity}
\eeq
in standard notation. Unlike the neutrino operators mentioned above,
these operators violate $B-L$ in one unit (rather than 2).
It is well known that the standard R-parity of the MSSM may be identified with a
$Z_2$ subgroup of $U(1)_{B-L}$, so  these terms are odd under R-parity.
The simultaneous presence of all these couplings is phenomenologically unacceptable. Indeed, the third coupling
violates baryon number,  and the other three violate lepton number. 
Together they lead to proton decay at an unacceptably large rate.
 On the other hand couplings violating {\it either } $B$ {\it  or }  $L$ are phenomenologically
allowed.

It is an interesting question whether any of these operators may be 
induced by string instanton effects. A first point to note is that 
instantons with $Sp(2)$ Chan-Paton symmetry can never generate operators of this 
type. The reason is that all charged  zero modes will necessarily come
in $Sp(2)$ doublets and hence the charged operators induced will
always involve an even number of charged $D=4$ fields and 
R-parity is automatically preserved. 
On the other hand $O(1)$ and $U(1)$ instantons may generate
R-parity violating operators. In particular, the $LH$ bilinear 
is essentially the square root of the Weinberg operator, and
may be induced if a $U(1)$ or $O(1)$  instanton  $M$  exists
with   
\beq
   { I_{M{\bf c}} = -1} \ ;\  {I_{M{\bf d}}=1 }
\ \ {\rm or}  \ \
\  {I_{M{\bf c}'} =1} \ ;\  {I_{M{\bf d}'}=-1 } \ .
\eeq
(in the $O(1)$ case the second option is not independent from the first).
Again, if the appropriate disk  couplings are  non-vanishing  a 
term  with $\mu_L ^a  \sim M_s \exp(-S_{ins})$ is generated.
The rest of the operators  in $W_{R_p}$ may  also be generated.
Possible instanton zero modes which may induce them are  
shown in table \ref{tableoperators}. 
For example, the $QDL$  operator may be induced if a $U(1)$  instanton $M$ 
with intersection numbers 
\beq
    I_{M{\bf b}} = -1 \ ;\  I_{M{\bf c}'}=1  \ ;\ I_{Md}=1
\eeq
is present and in addition couplings 
\beq
L_{disk}\ \propto \ c_{ab} \ (\alpha   (U^aQ^b_j) \gamma ^j)  \ 
+\  c_a' (\beta L^a_j\gamma_j ) 
\eeq
exist. Here $\alpha , \beta , \gamma $ are zero modes corresponding 
to $(Mc'), (Md)$ and $(bM)$ intersections and $a,b(j)$  are flavor($SU(2)_L$) indices.
Analogous trilinear or quartic disk amplitudes involving two charged zero modes
should exist to generate the rest of the  R-parity violating amplitudes 
in table \ref{tableoperators}.

\medskip

\begin{table}[htb] \footnotesize
\renewcommand{\arraystretch}{0.75}
\begin{center}
\begin{tabular}{|c||c|c|c|c|c|c|c|}
\hline $D=4$ Operator &
 $I_{M{\bf a}}$  &  $I_{M{\bf a}'}$ &  $I_{M{\bf b}}$  &  $I_{M{\bf c}}$ &
$I_{M{\bf c}'}$ &  $I_{M{\bf d}}$    & $I_{M{\bf d}'}$
\\
\hline\hline   $\nu_R\nu_R$ &  0  &  0   &  0   & 2  & 0 & -2   & 0\\
\hline $L{\bar H}L{\bar H}$  & 0   &  0   &  0    & -2  & 0 &  2  & 0
\\
\hline\hline  $L{\bar H}$  &  0    &  0  &  0    & -1  & 0 & 1    & 0
\\
\hline $QDL$  &    0    &  0   &  -1     & 0  & 1 &   1   & 0\\
\hline $UDD$  & -1   &  0   &  0   &  1  & 2   &  0   & 0 \\
\hline $LLE$  &  0    &  0   &  -1   & 0   & 1  & 1  & 0 \\
\hline \hline
 $QQQL$ &   1    &  0   &  -2      & 0  & 0 & 1   & 0 \\
\hline $UUDE$   &  -1    &  0   &   0  & 2  &  2  & -1    &  0  \\
\hline
\end{tabular}
\end{center}
\caption{\small  Zero modes required to generate Lepton/Baryon-number violating
superpotential operators. $Sp(2)$ instantons cannot give rise to R-parity violating 
operators whereas $O(1)$,$U(1)$ instantons may in principle contribute to all of them.
In the case of $U(1)$ instantons there are additional zero mode possibilities
which are obtained by exchanging $I_{M{\bf x}}\leftrightarrow - I_{M{\bf x}'}$.}
\label{tableoperators}
\end{table}

\subsubsection{\bf Dimension 5 proton decay operators }

There are also superpotential dimension-5 operators  violating
$B$ and $L$ which may be constructed from the 
MSSM matter superfields. Indeed the dimension  5 
operators
\beq 
(\frac {1}{M})\, QQQL \ \ \ ;\ \ \ (\frac {1}{M}) \, UUDE
\eeq
are in fact  the leading source of proton decay in 
SUSY GUT models with R-parity. Unlike the other operators considered here these ones 
preserve $B-L$ (hence R-parity) but not $B+L$.
These operators do not contribute 
directly to a proton decay but need to be 'dressed' by a one loop
exchange of some fermionic SUSY particle. This makes that, 
even although   they are suppressed only by one 
power of the relevant fundamental scale, the loop factor and
the corresponding couplings make the overall rate 
in SUSY-GUTS  (barely) consistent with present experimental bounds 
for $M$ of order the GUT scale or larger.

These dimension 5 operators may also be induced in 
D-brane models of the class here considered by 
the presence of instantons with 
appropriate intersection numbers. For instance, the first operator may be induced
through $O(1)$ or $U(1)$  instantons $M$  with
\beq
I_{M{\bf b}}=I_{M{\bf b}'}\ =\ -2 \  \ ;\  \    I_{M{\bf a}} =  1  \ ;\   I_{M{\bf d}}= 1 
\eeq
Again $Sp(2)$ instantons cannot induced this operator, since the the $Ma$ intersection would yield 6 (rather than 3) colored fermionic zero modes.
The proton decay rate obtained from these operators depend on the 
ratio $\rm{exp}(-S_{ins})\times 1/M_s$. For $M_s$ of order  $10^{16}$ GeV, 
the rate is consistent with present bounds if $\rm{exp}(-S_{ins})$ 
provides  a suppression of a few orders of magnitude. On the other hand,
models with a low  string scale may be in danger unless the 
exponential suppression
is sufficiently large (or such particular instantons are absent).

\medskip

As a general conclusion,  these phenomenological aspects of instanton induced
operators very much depend on the  action of the instanton, e.g.
the volume of the wrapped D2-instanton in the intersecting 
D-brane constructions. In any event it is clear that the
instantons here considered may indeed induce proton decay at 
a model-dependent rate. However in certain models R-parity will be
preserved  and prevent too rapid proton decay. 
Indeed, this is what we find in our instanton search in Gepner orientifolds.
As we said $Sp(2)$ instantons automatically preserve R-parity.
More generally,  models that violate R-parity are rare, and the corresponding instantons 
actually generate very high dimensional operators, so R-parity breaking effects seems quite suppressed. In fact in our search within MSSM-like models in Gepner model orientifolds we do not find instantons with just the correct charged zero modes 
to generate the low dimensional couplings discussed above. So, at least within our class of RCFT constructions, R-parity preservation is quite a common feature.

\section{CFT orientifolds}
\label{gepners}

In this section we describe the 4d string models we consider, namely orientifolds of type IIB 
Gepner model compactifications. This is a very large class, on which one can carry out 
large scans for certain desired properties. And moreover 
at present the only known  class of (SUSY) models with massive $B-L$.

\subsection{Construction of the models}
\label{gepnermodels}

In general, RCFT orientifolds are orientifold projections of closed
string theories constructed using rational conformal field theory. Although
this includes in principle rational tori and orbifolds, the real interest lies
in cases where the two-dimensional CFT is interacting, because such theories are
hard to access by other methods. A disadvantage of the use of RCFT is that this method
is algebraic, and not geometric in nature, so that one cannot easily explore small deformations
of a certain string theory. It is best thought of as a rational scan of moduli spaces.

The most easily accessible examples are the orientifolds of tensor products of minimal
$N=2$ conformal field theories (``Gepner models") forming a type IIB closed string theory.
During the last decade, examples in this class have been studied  by many authors
(see \cite{Angelantonj:1996mw}\cite{Blumenhagen:1998tj}\cite{Aldazabal:2003ub}\cite{Brunner:2004zd}\cite{Blumenhagen:2004cg}\cite{Dijkstra:2004ym}\cite{Aldazabal:2004by}\cite{Aldazabal:2006nz}), and searched systematically in \cite{schell1} and \cite{schell2}.
Although the Gepner models form only a small subset of RCFT's, they already offer a large
number of possibilities. The total number of tensor products with the required central charge
$c=9$ is 168. On top of this, one can choose a large number of distinct modular invariant partition
functions on the torus. The orientifold formalism is not available for all of them, but it has been
completely worked out \cite{Fuchs:2000cm} for all simple current invariants (based on the charge conjugation invariant).
This yields a total of 5403 distinct MIPFs. On top of this, we may choose various orientifold projections.
Here the only known possibilities are a class of simple-current based choices \cite{prss2}\cite{prss3}\cite{Huiszoon:1999xq}\cite{Huiszoon:2000ge}. This then yields a
total of 49304 orientifolds.

For each orientifold choice, the full open string partition function is
\beqa
\label{Spectrum}
\frac12 \left( \sum_{a,b,i} N_a N_b A^i_{ab} \chi_i (\frac{\tau}{2}) +
\sum_{a,i} N_a M^i_{a} \hat\chi_i (\frac{\tau}{2} + \frac12) \right)
\eeqa
Here $A^i_{ab}$ are the annulus coefficients, $M^i_a$ the Moebius coefficients, $N_a$ the
Chan-Paton multiplicities and $\chi(\tau)$ are the closed string characters, and $\hat\chi_i(\tau)=T^{-1/2}\chi_i(\tau)$.
The set of integers $i$ is simply the set of primary fields of the closed string CFT, and depends only on the tensor product.
The integers $a,b$ are the boundary labels; this set depends on the MIPF. Our notation
and labelling conventions for these CFT quantities are explained in Appendix A.
The integers $A^i_{ab}$ and $M^i_a$ depend in addition also on the orientifold choice; in the case of $A^i_{ab}$ the latter
dependence is very simple: all distinct annuli can be written as $A^{\Omega,i}_{ab} = \sum_c  A^{i~~c}_{a} C^{\Omega}_{cb}$, where
$\Omega$ is the orientifold choice (which we usually do not specify explicitly) and $C^{\Omega}_{cb}$ is the boundary conjugation
matrix, which acts as an involution on the set of boundaries.

Suppressing some details (which can be found in \cite{Fuchs:2000cm}) we may write these integers as
\beqa
A^{\Omega,i}_{ab} = \sum_{m,J,K} { S^i_m R_{a,(m,J)} g^{\Omega,m}_{JK} R_{b,(m,K)} \over S_{0m} }
\eeqa
\beqa
M^{\Omega,i}_{a} = \sum_{m,J,K} { P^i_m R_{a,(m,J)} g^{\Omega,m}_{JK} U^{\Omega}_{(m,K)} \over S_{0m} }
\eeqa
Here $m$ is the label of an Ishibashi-state (the set of states that propagates in the transverse
(or closed string) channel of the the annulus or Moebius diagrams). It is a subset of the set of
closed string labels $i$, but in general there are degeneracies, so that more than one distinct
Ishibashi state belongs to a given closed string label. These degeneracies are distinguished by the
labels $J,K$ (see Appendix A).
The complex numbers $R$ and $U$ are respectively the boundary and crosscap coefficients. Note
that the latter depend on the orientifold choice, but the former do not. The only dependence of the annulus
coefficients on the orientifold choice is through the Ishibashi metric $g^{\Omega}_{JK}$, which is a matrix
on each Ishibashi degeneracy space, and which can be a sign if there are no degeneracies. Finally, the matrix $P$
is given by $P=\sqrt{T}S T^2 S \sqrt{T}$, where $S$ and $T$ are the generators of the modular group
of the torus. Similar expressions exist for the Klein bottle multiplicities defining the unoriented
 closed sector, but they will not be needed in this paper.

The boundary labels $a,b,\ldots$ refer to all boundaries that respect the bulk symmetries of the CFT. This
includes the individual $N=2$ chiral algebras of the factors in the tensor product, the {\it alignment currents}\footnote{These
are spin-3 currents consisting of products of the world-sheet supercurrents of the factors in the tensor product, including the
NSR space-time factor.}
 that
ensure the proper definition of world-sheet supersymmetry and the space-time supersymmetry generator that
imposes a generalized GSO-projection on the spectrum. The latter implies that all characters $\chi_i$ respect
(at least) $N=1$ space-time supersymmetry. By construction, the boundary states are then supersymmetric as well.
Both conditions (boundary and bulk space-time supersymmetry) can in principle be relaxed within the formalism, but
this leads to a much larger set of bulk and boundary states.
The precise labelling of the boundaries is explained in Appendix A and involves
a subset of the closed string labels $i$ and a degeneracy label, distinct from the one used for the Ishibashi states.
The set of boundary labels is complete in the sense of \cite{prss3}. This means that no additional boundary states exist
that respect all the aforementioned symmetries. It also means that the matrices $R$ are square matrices (although
their rows and columns are defined in terms of different index sets). It is in principle possible to write down
additional boundary states that break some of the world-sheet symmetries. This is an important possibility to keep
in mind, but we will not consider it here.

The massless spectrum is obtained by restricting the characters $\chi_{i}$ to massless states. Since the characters
are supersymmetric those massless states are either vector multiplets or chiral multiplets. The latter can be
restricted to one chirality ({\it e.g} left-handed); the other choice merely produces the CPT conjugates. Boundaries
are called real if $a=a'$, where the conjugate boundary $a'$ is defined by $C^{\Omega}_{a,a'}=1$, and complex otherwise. The Chan-Paton multiplicities
$N_a$ give rise to gauge groups $U(N_a)$ for complex boundaries and $SO(N_a)$ or $Sp(N_a)$ for real ones. In the
latter case $N_a$ must be even. To count bi-fundamentals we define
\beqa
L_{ab} \equiv \sum_{i}  A^i_{ab}\chi_i (\frac{\tau}{2})_{{\rm massless}, L} \ .
\eeqa
Note that because of the factor $\frac12$ in (\ref{Spectrum}) and the fact that $L_{ab}$ is symmetric, the value of
$L_{ab}$ is indeed precisely the number of bi-fundamentals in the representation $(N_a,N_b)$.
It
is convenient to introduce the intersection matrix\footnote{Note that $L_{ab}$ is a symmetric matrix giving the number of chiral multiplets in the $(\fund_a,\fund_b)$
bi-fundamental. This is a natural quantity in unoriented CFT's, where a symmetric definition for the annulus amplitude exists. In oriented CFT the annulus is, in general, not  symmetric, but on the other hand it is possible to choose the branes in such a way that only $(\fund,\antifund)$ bi-fundamentals appear. This has become the customary way of counting states in the intersecting brane literature, even for orientifold models. The quantity $I_{ab}$ is defined in such a way that it is anti-symmetric in $a$ and $b$. This is why boundary conjugations appear in the
 right hand side. This has the additional advantage of making $I$ a more familiar quantity for readers used to the standard intersection brane conventions.}
 \beqa
\label{ChirInt}
I_{ab} \equiv L_{ab'} - L_{a'b} \ ,
\eeqa
which is manifestly antisymmetric in $a$ and $b$.
Note that for a pair of complex boundaries $a,b$ with conjugates $a',b'$ one can define four quantities
that are relevant for the massless spectrum, two of which are chiral, namely
$I_{ab}$ and $I_{ab'}$.

It is often convenient to associate a geometric picture to these integers. Thus we will often refer to the boundary
labels and their multiplicities as ``stacks of branes", and view the integers $I_{ab}$ as brane intersection numbers.
This is only done for convenience and
does not imply a concrete brane realization; indeed, it does not make sense to say that a given boundary label
corresponds to a D$p$-brane for some give $p$. Such an interpretation might be valid in a large radius limit, assuming
such a limit exists.

In general, for a choice of Chan-Paton multiplicities $N_a$ there will be tadpoles in the one-point closed string
amplitudes on the disk and the crosscap. These have to be cancelled in order to make the theory consistent (since we
work with supersymmetric strings we do not have the option of cancelling RR and NS-NS tadpoles separately). This
leads to a condition on the Chan-Paton multiplicities:
\beqa
\label{TadpoleCond}
\sum_a N_a R_{a,(m,J)} = 4 \eta_m U_{m,J}
\eeqa
where $\eta_0=1$ and all other $\eta$'s are $-1$; there is such a condition for any Ishibashi label $(m,J)$ that leads
to a massless scalar in the transverse channel. The one for $m=0$ (which is non-degenerate) is the dilaton tadpole condition.
It has the special feature that all coefficients $R_{a0}$ are real and positive.
The crosscap coefficient $U_{0}$ is also real and can be chosen
positive (in the CFT both signs are acceptable). If $U_{0} \not= 0$ (\ref{TadpoleCond}) limits the Chan-Paton multiplicities;
if $U_{0} = 0$ the only solution is $N_a=0$ for all $a$, which rules out any realization of the Standard Model.
This reduces the number of usable orientifolds to 33012.

Tadpole cancellation condition implies cancellation of RR-charges coupling to long-range fields, and
absence of local anomalies. There is a second condition
that has to be taken into account, which has to do with ${\bf Z}_2$ charges that do not
couple to long-range fields, usually referred to as ``K-theory charges" in geometric constructions.
Uncancelled K-theory charges may lead to global anomalies in symplectic factors of the gauge group. But even
if this symptom is absent, the disease may still exist. A much more general way to probe for uncancelled K-theory charges is
to require the absence of global anomalies not only in the Chan-Paton gauge group but also on all symplectic
brane-anti-brane pairs that can be added to it as ``probe-branes" \cite{Uranga:2000xp}. Presently this is the most general
constraint that be imposed in these models, but it is not known if additional ones are required. This probe brane
constraint leads to a large number of mod-2 constraint and is potentially very restrictive, but almost harmless in practice \cite{Gato-Rivera:2005qd}.
It is satisfied by all models we consider in the present paper.

\subsection{Search for SM-like models}
\label{gepnersm}

The complete set of solutions to these conditions is finite but huge, but the vast majority is of no phenomenological
interest. In the last few years systematic searches have been carried out for models that contain the Standard Model.
The models that were considered have the
property that the set of Chan-Paton labels can be split into two subsets, the
observable and the hidden sector.
The former has been limited, for practical reasons, to at most four complex brane stacks, required to contain the Standard Model
gauge group and the right intersections to yield three families of quarks and leptons, plus (in general) some non-chiral
(vector-like) additional matter. The hidden sector is only constrained by the requirement that there be no net number of chiral multiplets charged under both the observable and hidden sector,
and by practical computational limitations. The main purpose of the hidden sector in these models is to provide variables that
can be used to satisfy the tadpole and global anomaly conditions, since the multiplicities in the observable sector are already fixed. In some cases the observable
sector already satisfies the constraints by itself, and there is no hidden sector.

The observable sector can be realized in many different ways if one only imposes the constraint
that the standard model should be contained in it. These possibilities were recently explored
in \cite{schell2}. We will focus on the realization described in Section \ref{mssm}, first considered in \cite{imr}. There are four stacks, namely {\bf a} (containing QCD and baryon number as $U(3)$), ${\bf b}$ (containing electroweak $SU(2)$ embedded as $U(2)$ or $Sp(2)$), ${\bf c}$  (providing a $U(1)$ or an $O(2)$ factor\footnote{In \cite{schell1} also $Sp(2)$ was considered, but this requires an additional Higgs mechanism.}, and ${\bf d}$ (providing another $U(1)$ factor).

The standard model hypercharge generator is , defined in (\ref{Ydef}):
\beqa
Y= \frac16 Q_{\bf a} - \frac12 Q_{\bf c} - \frac12 Q_{\bf d}
\eeqa
where $Q_{\bf x}$ denotes the generator of the $U(1)$ of brane stack ${\bf x}$; in case the Chan-Paton factor of brane {\bf c} is $O(2)$ one should use the properly normalized $O(2)$ generator. In addition to $Y$ these models have two or three additional $U(1)$ gauge symmetries (the latter case if electroweak $SU(2)$ arises from $U(2)$). These (except the combination $B-L$) are anomalous, with anomaly cancelled by the Green-Schwarz mechanism, implying the existence of a $B\wedge F$ coupling making them massive. In fact, as already mentioned, such St\"uckelberg couplings may be present for non-anomalous $U(1)$'s as well. We
are interested in models where the hypercharge gauge boson does not have such couplings (otherwise the model would be phenomenologically unacceptable), but where the $B-L$ gauge boson is massive by such couplings (both in order that the gauge group reduces to the SM one, and that neutrino Majorana masses may be induced by string instantons, as discussed in previous sections).

The combined requirements of having a massive $B-L$ and a massless $Y$ turn out to be difficult to satisfy.
In fact, if the group on brane ${\bf c}$ is $O(2)$ they are impossible to satisfy simultaneously, because the $O(2)$ component of the vector boson does not couple to any axions, and hence the $B-L$ and $Y$ bosons have the same mass.
But even in models with a $U(1)$ group on brane {\bf c} it happens rather rarely that both constraints are satisfied simultaneously, at least in the searches that have been done so far.

We will make use here of the data presented in \cite{schell1,schell2}, 
which are available in slightly improved
form on the website \href{http://www.nikhef.nl/~t58/filtersols.php}{www.nikhef.nl/$\sim$t58/filtersols.php}. This database consist of
211634 distinct spectra. Here ``distinct" means that they are physically different
for a given MIPF\footnote{Rare cases of identical spectra and couplings originating  from different MIPFs are treated as distinct.}
if the hidden sector
is ignored. Hence the differences can be the number of vector-like states of various kinds or the dilaton couplings of
branes {\bf a, b, c, d}. Geometrically, these spectra may originate from the same moduli space, but then in any case
from different points on this moduli space. The improvements in comparison with the data presented in \cite{schell1} consist of
taking into account the full global anomaly conditions from probe branes. In some cases this required nothing more than
checking these conditions for an existing solution of the tadpole conditions, but in other cases a new solution had to be
found. As a result, a few models disappeared from the original database, but due to improved algorithms a few new ones could be
added. The net result is some small but inconsequential changes in the total number of models of various kinds. The numbers we
will mention below are based on the improved database.

The total number of models in that database with a Chan-Paton group $U(3)\times Sp(2) \times U(1) \times U(1)$ is 10587.
Of these, 391 (about $4\%$) have a massive $B-L$ vector boson. For $U(3)\times U(2) \times U(1) \times U(1)$ these
numbers are, respectively, 51 and 0. Hence no examples of the latter type were found, although they were found with
1,2 and 4 families (in a limited search), in a few percent of the total number of models. It seems therefore reasonable
to expect that $U(3)\times U(2) \times U(1) \times U(1)$ with massive $B-L$ do exist, and that their absence is just
a matter of statistics. Just for comparison, the total number of $U(3)\times Sp(2) \times O(2) \times U(1)$ models is 56627.

\section{Fermion zero modes for instantons on RCFT's}
\label{themess}

In this section we discuss D-brane instantons for general compactifications, including abstract CFT ones. We also provide the spectrum of zero modes on an instanton brane, using the information about their internal structure i.e. in the compactified dimension in geometric models, or of the internal CFT in more abstract setups like in previous section. We will be interested in the latter case.

A first question that should be addressed is what this internal structure is. For instance, in type IIA geometric compactifications, it corresponds to a supersymmetric (i.e. special lagrangian) 3-cycle. Notice that these are the same kind of 3-cycles already used to wrap the D6-branes that give rise to the 4d gauge symmetry of such models. For general CFT's, D-branes are described as boundary states. To describe instantons, one can simply use the same boundary state of the internal CFT to describe the 4d space-filling branes present in the model and the instanton branes. The only difference is that boundaries satisfy Neumann conditions in the 4d space-filling case, and Dirichlet in the instanton case. This exploits the fact that whenever a boundary state of the internal CFT, and with Neumann boundary conditions in the 4d space is an acceptable state of the full CFT, the same boundary state of the internal CFT, combined with Dirichlet boundary conditions in the 4d space also gives an acceptable state of the full CFT. 
For geometric compactifications this is related to Bott periodicity of the K-theory classes associated to the D-brane charges, but it is possible to show it in general.

Since instanton D-branes can thus be naturally associated to the boundary 
states of 4d space-filling branes, it is convenient to express the 
spectrum of zero modes of the former in terms of the massless states of the latter. This is particularly useful, since the computation of the spectra on 4d space-filling branes for Gepner model orientifolds has already been described (although the arguments below are valid also for geometric compactifications).  Hence, let us denote by ${\cal M}$ a 4d space-filling brane associated with the same boundary state of the internal CFT as the instanton brane $M$ of interest. Note that the 4d space-filling brane ${\cal M}$ is an auxiliary tool, and need not be (and, for our instantons of interest, will not be) one of the 4d space-filling branes present in the model.

\medskip

{\bf `Real' brane instantons}

Let us first consider the case of `real' brane instantons. Consider a set of $m$ 4d space-filling branes ${\cal M}$, and focus first on the massless spectrum in the ${\cal M}{\cal M}$ sector.
Before the orientifold projection, it leads to a universal 4d $N=1$ $U(m)$ vector multiplet, and a number $L_{{\cal M}{\cal M}}$ 
of adjoint chiral multiplets. The orientifold operation maps this sector to itself, acting on the Chan-Paton with a matrix $\gamma_{\Omega, \cal M}$. This matrix satisfies 
\beqa
\gamma_{\Omega,{\cal M}}^T \gamma_{\Omega,{\cal M}}^{-1}\, =\, \pm \id_m
\label{sospfill}
\eeqa
The two possibilities can be chosen to correspond to $\gamma_{\Omega,{\cal M}}=\id_m$ or $\gamma_{\Omega,{\cal M}}=\epsilon_m$, with $\epsilon_m=\pmatrix{ 0 & \id_r \cr -\id_r & 0}$, and $m=2r$ hence necessarily even in the latter case. They correspond to the $SO$ and $Sp$ projections, respectively. 

The orientifold projection on the $N=1$ vector multiplet Chan-Paton 
matrices is given by 
\beqa
\lambda \, =\, -\gamma_{\Omega,{\cal M}} \, \lambda^T\, \gamma_{\Omega,{\cal M}}^{-1}
\label{proyvect}
\eeqa
and leads to $SO(m)$ or $Sp(m)$ vector multiplets for the $SO$ or $Sp$ projection (hence the name). Concerning the $N=1$ chiral multiplets, they fall in two classes of $p_-$, $p_+$ (with $p_-+p_+=L_{{\cal M}{\cal M}}$) which  suffer the projections
\beqa
\lambda \, =\, \pm \gamma_{\Omega,{\cal M}} \, \lambda^T\, \gamma_{\Omega,{\cal M}}^{-1}
\label{proychir}
\eeqa
For the $SO$ projection, this leads to $p_+$, $p_-$ chiral multiplets in the $\Ysymm$, $\Yasymm$ representation. For the $Sp$ projection, there are $p_+$, $p_-$ chiral multiplets in the $\Yasymm$,   $\Ysymm$ representation.

 The sectors ${\cal M}a$ (where $a$ is a 4d space-filling branes present in the model) are mapped to sectors ${\cal M}a'$, so it is enough to focus on the former. After the orientifold projection one gets
 $L_{{\cal M}a}$, $L_{{\cal M}a'}$ chiral multiplets in the $(\fund_{\cal M}, \antifund_a)$,  $(\fund_{\cal M}, \fund_a)$.
 
Let us now obtain the zero modes for a set of $m$ instanton branes $M$ in 
terms of the above spectrum. The $MM$ sector is closely related to the 
${\cal M}{\cal M}$ sector, by changing the NN boundary conditions 
in 4d spacetime to DD boundary conditions (which can be done in a covariant
formalism, but not in the light-cone gauge). Before the orientifold 
projection, one obtains the same set of states (since moddings for NN and 
DD boundary conditions are identical, both in the NS and R sector), but 
with different world-volume interpretation. Also, the change in boundary conditions 
implies that some polarization states which are unphysical for the 4d spacefilling 
brane are physical in the instanton brane.
Hence, the $U(m)$ gauge bosons on the 4d space-filling brane ${\cal M}$ correspond to four 
adjoint real scalars in the instanton brane $M$. Similarly, the 4d spinors in ${\cal 
M}$, correspond to four fermion zero modes on $M$, transforming as two spinors of opposite chiralities
$\theta^\alpha$, ${\tilde \theta}_{\dot \alpha}$
of the $SO(4)$ rotation group in transverse space. The orientifold projection maps the 
$MM$ sector to itself, acting on Chan-Paton indices with a matrix 
$\gamma_{\Omega, M}$. In close analogy with the argument in \cite{gp} for 
the familiar D5-D9-brane system in type I (see 
\cite{Pradisi:1988xd,Witten:1995gx} for related derivations), one can show that the condition 
(\ref{sospfill}) flips sign upon changing four NN boundary conditions to 
DD, hence
\beqa
\gamma_{\Omega,M}^T \gamma_{\Omega,M}^{-1}\, =\, \mp \id_m
\label{sospinst}
\eeqa
Namely, the instanton brane has $Sp(m)$ gauge group when the 4d space-filling brane (with same internal boundary state) has gauge group $O(m)$, and vice-versa. We still refer to these projections as $SO$ and $Sp$, hoping no confusion arises. 
Note that, as mentioned in Section \ref{zeromode}, although there are no gauge 
bosons in $0+0$ dimensions, the gauge group is present on the instantons 
in that it acts on open string endpoints.

Let us consider the effect of the orientifold projection on the $MM$ states, as compared with the effect on ${\cal M}{\cal M}$ states. Again, following arguments familiar in the D5-D9 brane system in type I, one can show that the signs in conditions like (\ref{proyvect}), (\ref{proychir}) remain unchanged upon changing four NN dimensions to DD, except for bosonic modes polarized along the directions longitudinal to these four dimensions (and for fermions related to them by the unbroken susy of the total system). To be concrete, considering the four $MM$ adjoint bosons, and two $MM$ adjoint fermions $\theta^\alpha$ associated to the universal ${\cal M}{\cal M}$ vector multiplets, they suffer the projection
\beqa
\lambda \, =\, + \gamma_{\Omega,M} \, \lambda^T\, \gamma_{M}^{-1} 
\eeqa
Hence they transform in the $\Yasymm$ of $Sp(m)$ for the $SO$ projection, and in the $\Ysymm$ of $SO(m)$ for the $Sp$ projection. 
On the other hand, for the two fermion zero modes $\tilde\theta_{\dot\alpha}$, the projection is
\beqa
\lambda \, =\, - \gamma_{\Omega,M} \, \lambda^T\, \gamma_{M}^{-1} 
\eeqa
and leads to two fermion zero modes in the $\Ysymm$ of $Sp(m)$ for the $SO$ projection, and in the $\Yasymm$ of $SO(m)$ for the $Sp$ projection.

This implies that in order to obtain two fermion zero modes from this universal multiplet, in order to generate a superpotential, one should consider instantons with orthogonal gauge group and multiplicity one ($O(1)$ instantons). For instantons with symplectic gauge group and multiplicity two ($Sp(2)$ instantons), there are two additional fermion zero modes in the triplet representation. As mentioned, we will continue to consider such instantons in our relaxed scan.
Multiple instantons, i.e. boundary states with higher multiplicity, lead to a larger amount of additional fermion zero modes (due to the larger gauge representations for the fermions), and do not contribute to superpotentials; we will not consider such cases even in relaxed scans, since they also very often lead to too many charged fermion zero modes and cannot contribute to the operators of interest (except possibly for $O(2)$ and $U(2)$ instantons with low intersections, which are kept in our scan as a curiosity).
 
\medskip

Similarly, for the $p_{\pm}$ sets of $MM$ scalars and fermions associated to the ${\cal M}{\cal M}$ 4d chiral multiplets, the projection is
\beqa
\lambda \, =\, \pm \gamma_{\Omega, M} \, \lambda^T\, \gamma_{\Omega, M}^{-1}
\eeqa
with the same sign choice as in (\ref{proychir}). The different structure of $\gamma_{\Omega}$ implies that, for the $SO$ projection we get $p_+$, $p_-$ sets of scalars and fermions in the $\Yasymm$, $\Ysymm$, while for the $Sp$ projection there are $p_+$, $p_-$ sets of scalars and fermions in the $\Ysymm$, $\Yasymm$.

This concludes the discussion of the $MM$ sector. Let us not consider the $Ma$ sectors, from the information from the ${\cal M}a$ sectors. Notice that this implies changing four NN boundary conditions to DN, which have different moddings. Hence the states are different in both situations, but the information on the multiplicities is preserved. Specifically, in the NS sector the DN boundary condition introduce an additional vacuum energy which generically makes all states massive. Hence there are no massless scalar zero modes in generic ${\cal M}a$ sectors. In the R sector, the change in the moddings reduces the dimension of the massless ground state, leading to a single (chiral) fermionic degree of freedom. Since the orientifold action maps the $Ma$ sector to $Ma'$ sectors, there are no subtleties in the orientifold projection.
The end result is $L_{{\cal M}a}$, $L_{{\cal M}a'}$ fermion zero modes in the $(\fund_{\cal M}, \fund_a)$, $(\fund_{\cal M}, \antifund_a)$. The net number of chiral fermion zero modes in the 
$(\fund_{\cal M}, \antifund_a)$ is given by $I_{{\cal M}a}=L_{{\cal M}a'}-L_{{\cal M}a}$, i.e. the net number of chiral multiplets in the related ${\cal M}a$ sector.

The results for orientifold projections for real branes are shown in table \ref{realproy}.

\begin{table}[h!]
\begin{center}
~~~~~~~~~~~~~~~~\begin{tabular}{|c|c|c|c|c|} \hline
 Proj. & Multiplet in ${\cal M}$ & ${\cal M}$ (before orient.) & ${\cal M}$ (after orient.) & $M$ (after orient.)\\ \hline\hline
SO & $N=1$ vect. mult. & $U(m)$ & $O(m)$ & $Sp(m)$ \\
& & & & $2\,\Yasymm_{\, f}+ 2\,\Ysymm_{\, f}+4\, \Yasymm_{\, b}$ \\ \hline
& $N=1$ ch. mult. & $(p_++p_-)\, \Ad$ & $p_+ \, \Ysymm \, +\, p_-\, \Yasymm$ & $2p_+\, (\,\Yasymm_{\, f}\, +\, \Yasymm_{\, b}\, )\, +$ \\ 
& & & & $2p_-\, (\,\Ysymm_{\, f}\, +\, \Ysymm_{\, b}\, )$\\ \hline\hline
Sp & $N=1$ vect. mult. & $U(m)$ & $Sp(m)$ & $O(m)$ \\
& & & & $2\,\Ysymm_{\, f}+2\,\Yasymm_{\, f}+4\, \Ysymm_{\, b}$ \\ \hline
& $N=1$ ch. mult. & $(p_++p_-)\, \Ad$ & $p_+ \, \Yasymm \, +\, p_-\, \Ysymm$ & $2p_+\, (\,\Ysymm_{\, f}\, +\, \Ysymm_{\, b}\, )\, +$ \\
& & & & $2p_-\, (\,\Yasymm_{\, f}\, +\, \Yasymm_{\, b}\, )$\\ \hline\hline
Any & $N=1$ ch. mult. & $L_{{\cal M}a'} (\fund_{\cal M},\antifund_a) +$ &  $L_{{\cal M}a'} (\fund_{\cal M},\antifund_a) +$ & $L_{Ma'} (\fund_M,\antifund_a)_{\, f}$\\ 
 & & $L_{{\cal M}a} (\fund_{\cal M},\antifund_{a})$ & $L_{{\cal M}a} (\fund_{\cal M},\fund_{a})$ & $L_{Ma} (\fund_M,\fund_{a})_{\, f}$ \\
 \cline{4-5} 
& & & net $I_{{\cal M}a} (\fund_{\cal M},\antifund_a)$ & net $I_{Ma} (\fund_M,\antifund_a)_{\, f}$\\ \hline 
\end{tabular}
\caption{\small Orientifold projection for real branes: Massless modes of the 4d space-filling branes ${\cal M}$ (before and after the orientifold projection) and zero modes on the instanton branes $M$ (denoted with sub-indices $b,f$ for bosonic and fermionic modes)}
\label{realproy}
\end{center}
\end{table}

\medskip

{\bf Complex brane instantons}

We now consider the case of complex brane instantons. The arguments are very similar, hence the discussion is more sketchy. Consider $m$ 4d spacefilling branes ${\cal M}$, associated to the internal boundary state of the instanton brane $M$ of interest. The ${\cal M}{\cal M}$ leads to a 4d $N=1$ $U(m)$ vector multiplet and a number $L_{{\cal M}{\cal M}'}$ of adjoint chiral multiplets. The orientifold action maps it to the ${\cal M}'{\cal M}'$ sector, hence we may keep just the former and impose no projection. The ${\cal M}{\cal M}'$ sector is mapped to itself under the orientifold projection. Denoting by $\gamma_{\Omega,{\cal M}}$ the action on Chan-Paton indices, the ${\cal M}{\cal M}'$  modes split into sets $L^{\pm}_{MM}$, $L^{\pm}_{M'M'}$, which suffer a projection
\beqa
\lambda \, =\, \pm \gamma_{\Omega,{\cal M}} \, \lambda^T\, \gamma_{\Omega,{\cal M}}^{-1}
\eeqa
leading, for $\gamma_{\Omega,{\cal M}}=\id_m$, to $L^{+}_{{\cal M}{\cal M}}$, $L^-_{{\cal M}{\cal M}}$ chiral multiplets in the  $\Ysymm$, $\Yasymm$, and $L^{+}_{{\cal M}'{\cal M}'}$, $L^-_{{\cal M}'{\cal M}'}$ chiral multiplets in the  $\bYsymm$, $\bYasymm$. The net number of chiral multiplets in the $\Ysymm$, $\Yasymm$ is $I^+_{{\cal M}{\cal M}'}=L^+_{{\cal M},{\cal M}}-L^+_{{\cal M}'{\cal M}'}$,  $I^-_{{\cal M}{\cal M}'}=L^-_{{\cal M},{\cal M}}-L^-_{{\cal M}'{\cal M}'}$. And oppositely for $\gamma_{\Omega,{\cal M}}=\epsilon_m$.

Finally, the ${\cal M}a$, ${\cal M}a'$ and related sectors lead, after the orientifold projection, to $L_{{\cal M}a'}$, $L_{{\cal M}a}$, $L_{{\cal M}'a'}$, $L_{{\cal M}'a}$ chiral multiplets in the $(\fund_{\cal M},\antifund_a)$, $(\fund_{\cal M}, \fund_a)$, $(\antifund_{\cal M},\antifund_a)$, $(\antifund_{\cal M},\fund_a)$. In order to simplify notation, we replace ${\cal M}\to M$ in these expressions in our discussions of instanton zero modes.

\medskip

Let us now consider $m$ brane instantons $M$ and compute their zero mode spectrum in terms of the above. In the $MM$ (and its image $M'M'$) sector there are four scalar modes and four fermions in the adjoint of the $U(m)$ gauge symmetry group; these are related to the 4d vector multiplet in the ${\cal M}{\cal M}$ sector. In addition, there are $L_{MM'}$ sets of scalars and fermions in the adjoint, related to the $L_{{\cal M}{\cal M'}}$ non-universal chiral multiplets in the ${\cal M}{\cal M}$ sector. The  $MM'$ sector is mapped to itself, and one has to impose the orientifold projection (recalling that the matrix $\gamma_{\Omega,M}$ differs from $\gamma_{\Omega,{\cal M}}$). For 
$\gamma_{\Omega,{\cal M}}=\id$, hence $\gamma_{\Omega,M}=\epsilon$, we obtain $L^{+}_{MM}$, $L^-_{MM}$ chiral multiplets in the  $\Yasymm$, $\Ysymm$, and $L^{+}_{M'M'}$, $L^-_{M'M'}$ chiral multiplets in the  $\bYasymm$, $\bYsymm$. The net number of chiral multiplets in the $\Yasymm$, $\Ysymm$ is $I^+_{MM'}=L^+_{MM}-L^+_{M'M'}$,  $I^-_{MM'}=L^-_{MM}-L^-_{M'M'}$. And oppositely for $\gamma_{\Omega,{\cal M}}=\epsilon$ hence $\gamma_{\Omega,M}=\id$.

In the $Ma$, $Ma'$ and related sectors, there are generically no bosonic zero modes, and there are
$L_{{\cal M}a}$, $L_{{\cal M}'a'}$, $L_{{\cal M}a'}$, $L_{{\cal M}'a}$ chiral fermion zero modes in the  $(\fund_M,\fund_a)$, $(\antifund_M,\antifund_a)$, $(\fund_M,\antifund_a)$, and $(\antifund_M,\fund_a)$ respectively. The net number of chiral fermion zero modes in the $(\fund_M,\antifund_a)$ and $(\fund_M, \fund_a)$ is given by $I_{{\cal M}a}=L_{{\cal M}a'}-L_{{\cal M}'a'}$ and $I_{{\cal M}a'}=L_{{\cal M}a}-L_{{\cal M}'a}$. In order to simplify notation, we replace ${\cal M}\to M$ in these expressions in our discussions of instanton zero modes.

The results for orientifold projections for real branes are shown in table \ref{cmplxproy}.

\begin{table}[h!]
\hspace{-3cm}
~~~~~~~~~~~~~~~~\begin{tabular}{|c|c|c|c|c|} \hline
 Proj. & Multiplet in ${\cal M}$ & ${\cal M}$ (before orient.) & ${\cal M}$ (after orient.) & $M$ (after orient.)\\ \hline\hline
 Any & $N=1$ vect. mult. & $U(m)\times U(m)'$ & $U(m)$ & $U(m)$ \\
& & & & $4\,\Ad_{\, f}+4\, \Ad_{\, b}$ \\ \hline
& $N=1$ ch. mult. & $p_{\rm adj}\, \Ad\, + p_{\rm adj} \Ad'$ & $p_{\rm adj} \Ad$ & $2p_{\rm adj}\, (\,\Ad_{\, f}\, +\, \Ad_{\, b}\, )$ \\  \hline\hline
$SO$ & $N=1$ ch.mult. & $L_{{\cal M}{\cal M}} (\fund_{\cal M},\fund_{\cal M'})$ &  $L^+_{\cal M\cal M}\, \Ysymm_{\cal M} +L^-_{\cal M\cal M}\, \Yasymm_{\cal M}$ &
$2L^+_{MM}\, \Yasymm_{\, b,f} +2L^-_{MM}\, \Ysymm_{\, b,f}$ \\ \hline
 &  & $L_{{\cal M'}{\cal M'}} (\antifund_{\cal M},\antifund_{\cal M'})$ &  $L^+_{\cal M'\cal M'}\, \bYsymm_{\cal M} +L^-_{\cal M'\cal M'}\, \bYasymm_{\cal M}$ &
$2L^+_{M'M'}\, \bYasymm_{\, b,f} +2L^-_{M'M'}\, \bYsymm_{\, b,f}$ \\  \hline\hline
$Sp$ & $N=1$ ch.mult. & $L_{{\cal M}{\cal M}} (\fund_{\cal M},\fund_{\cal M'})$ &  $L^+_{\cal M\cal M}\, \Yasymm_{\cal M} +L^-_{\cal M\cal M}\, \Ysymm_{\cal M}$ &
$2L^+_{MM}\, \Ysymm_{\, b,f} +2L^-_{MM}\, \Yasymm_{\, b,f}$ \\ \hline
 &  & $L_{{\cal M'}{\cal M'}} (\antifund_{\cal M},\antifund_{\cal M'})$ &  $L^+_{\cal M'\cal M'}\, \bYasymm_{\cal M} +L^-_{\cal M'\cal M'}\, \bYsymm_{\cal M}$ &
$L^+_{M'M'}\, \bYsymm_{\, b,f} +L^-_{M'M'}\, \bYasymm_{\, b,f}$ \\ \hline\hline
Any & $N=1$ ch. mult. & $L_{{\cal M}a'} (\fund_{\cal M},\antifund_a) +$ &  $L_{{\cal M}a'} (\fund_{\cal M},\antifund_a) $ & $L_{Ma'} (\fund_M,\antifund_a)_{\, f} $\\ 
 & & $\ldots$ & $L_{{\cal M}a} (\fund_{\cal M},\fund_{a})$ & $L_{Ma} (\fund_M,\fund_{a})_{\, f}$ \\ 
 & & $\ldots$ & $L_{{\cal M}'a'} (\antifund_{\cal M},\antifund_{a})$ & $L_{M'a'} (\antifund_M,\antifund_{a})_{\, f}$ \\ 
 & & $\ldots$ & $L_{{\cal M}'a} (\antifund_{\cal M},\fund_{a})$ & $L_{M'a} (\antifund_M,\fund_{a})_{\, f}$ \\ 
 \cline{4-5}
& & & net $I_{{\cal M}a} (\fund_{\cal M},\antifund_a)$ & net $I_{Ma} (\fund_M,\antifund_a)_{\, f}$\\
& & & net $I_{{\cal M}a'} (\fund_{\cal M},\fund_a)$ & net $I_{Ma'} (\fund_M,\fund_a)_{\, f}$\\ \hline 
 \hline
\end{tabular}
\caption{\small Orientifold projection for complex branes: Massless modes of the 4d space-filling branes ${\cal M}$ (before and after the orientifold projection) and zero modes on the instanton branes $M$ (denoted with sub-indices $b,f$ for bosonic and fermionic modes)}
\label{cmplxproy}
\end{table}

 \section{Search for $M$ instantons}
\label{scan}

In this section we perform a search of models which admit an instanton inducing 
a right-handed neutrino Majorana mass operator. Namely, for each model with 
the chiral content of the SM in the classification described in Section 
\ref{gepnersm}, we first scan over boundary states, searching for all  instantons with 
 the required uncharged and charged fermion zero mode structure  to yield neutrino masses.
We then relax our criteria a bit and allow for instantons with correct
charged zero mode structure but having extra non-chiral zero modes 
(both charged and uncharged). The idea is that these non-chiral 
zero modes could be lifted by diverse effects, as discussed.

It is important to recall that the cubic couplings between instanton zero
 modes and 4d chiral multiplets are difficult to compute in Gepner model
 orientifolds. Hence, we will simply assume that such couplings are
 non-zero if there is no symmetry forbidding them.

\subsection{The instanton scan}

Our detailed strategy will become clear along the description of the results.
Given a set of {\bf a,b,c,d} standard model branes, we must look for additional boundary states $M$ that satisfy the requirements of a $(B-L)$-violating instanton. From the internal CFT point of view this is just another boundary state, differing from 4d spacefilling branes only in the fully localized 4d spacetime structure. The minimal requirement for such a boundary state is $B-L$ violation, which means explicitly
\beqa
\label{AnomalyCond}
 I_{M{\bf a}}-I_{M{\bf a}'}-I_{M{\bf d}}+I_{M{\bf d}'} \not= 0
\eeqa
It is easy to see that the existence of such an instanton implies (and hence requires) the existence of a 
St\"uckelberg coupling making $B-L$ massive. To see this, 
consider adding to the Standard Model configuration a 4d
 spacefilling brane ${\cal M}$ (in fact used in Section \ref{themess})
 associated to the boundary state $M$ (RR tadpoles can be avoided by 
simultaneously including $M$ antibranes, which will not change
 the argument). The new sector in the chiral spectrum charged under the branes ${\cal M}$ can be obtained by reversing the argument in Section \ref{themess}, and is controlled by the intersection numbers of $M$.
From the above condition it follows that the complete system has mixed 
 $U(1)_{B-L} \times (G_{\cal M})^2$ anomalies, where $G_{\cal M}$ is the 
Chan-Paton-factor of brane ${\cal M}$. These anomalies are cancelled 
by a Green-Schwarz mechanism involving a $(B-L)$-axion bilinear coupling, which ends up giving a mass 
to 
$B-L$ via the St\"uckelberg mechanism. This coupling is in fact not sensitive to the presence of the brane ${\cal M}$, hence it must have been present already in the initial model (without ${\cal M}$).

Hence the existence of a boundary label $M$ that satisfies (\ref{AnomalyCond}) implies that $B-L$ is massive. Unfortunately the converse is not true: even if $B-L$ has a St\"uckelberg mass, this still does not imply the existence of suitable instantons 
satisfying (\ref{AnomalyCond})\footnote{From intuition in geometric compactifications, one expects 
that there may always exist a D-brane with the appropriate topological pairings,
 but there is no guarantee that there is a supersymmetric representative in that
 topological sector, and even less that  it  would have no additional fermion zero modes. 
Note also that even if such D-brane instantons exists, there is no guarantee 
that it will fall in the scan over RCFT boundary states.} Indeed, in several models
we found not a single boundary state satisfying (\ref{AnomalyCond}).

Note that, since hypercharge must be massless, one can use the reverse argument and obtain that
\beqa
I_{M{\bf a}}-I_{M{\bf a}'}-I_{M{\bf c}}+I_{M{\bf c}'}-I_{M{\bf d}}+I_{M{\bf d}'} = 0
\eeqa
in all models. 
We verified this for all models we considered as a check on the computations.

\smallskip

As already discussed in Section \ref{gepnersm}, in the search for SM constructions in Gepner orientifold, there are 391 models with massless hypercharge and massive $B-L$. In these models we found a total of 29680 instantons with $B-L$ violation, i.e. with intersection numbers satisfying (\ref{AnomalyCond}). Of course, in order to serve our purpose of generating a Majorana mass superpotential, the instantons have to satisfy some more conditions. Let us
consider them in order of importance, and start with the conditions on the net number of chiral fermion zero modes charged under the 4d observable sector.
Clearly we need $I_{M{\bf a}}=I_{M{\bf a}'}$ and $I_{M{\bf b}}=I_{M{\bf b}'}$. The latter condition is automatically satisfied in this case, because the {\it b}-brane is real in all 391 models.
The chiral conditions on the zero modes charged under the branes ${\bf c}$ and ${\bf d}$ are as in
\cite{iu}\footnote{Note that there is a sign change in
the contribution of the $U(1)_{\bf d}$ generator to $Y$ in comparison to \cite{iu}}
and are given in equations (\ref{InstInterOne}), (\ref{InstInterTwo}) (\ref{InstInterThree})
 of the present paper. 
These are the instantons of most interest, and on which we mainly focus.
 However, as discussed in Section \ref{otherbl}, other important $B$- and/or $L$-
  violating operators (such as the Weinberg operator or the $LH$ operator) can be generated 
by instantons with similar intersection numbers, up to a factor of 2 and a sign,
 see table \ref{tableoperators}. For this reason we also allow at this stage any
 instanton which has the correct number of charged zero modes to generate them.
Imposing these conditions reduces the number of candidate instantons potentially contributing to neutrino masses 
in any of the models to 1315. 

All instantons satisfying these requirements are summarized
in the table \ref{tbl:TableInstanton}. In columns 1,2 and 3 we list the tensor combination,
 MIPF and orientifold choice for which the model occurred.  The latter two numbers 
 codify simple current data that describe respectively a MIPF and an orientifold. MIPFs are
 in general defined by means of a subgroup $ {\cal H}$ of the simple current group ${\cal G}$,
 plus a certain matrix $X$ of rational numbers \cite{Kreuzer:1993tf}. Orientifolds are defined
 by a simple current and a set of signs \cite{Fuchs:2000cm}. In previous work \cite{schell1} we have
 enumerated these quantities (up to permutation symmetries) and assigned integer labels to them
 for future reference. We only refer to these numbers here, but further details are available upon
 request.
 Usually for each MIPF and orientifold which contains the standard model  there are
several choices {\bf a,b,c,d} for which it is obtained. 
For a given choice of tensor combination, MIPF and orientifold and SM branes there 
may be several instantons. For clarity we put all such instantons
 together in the information in table \ref{tbl:TableInstanton}. In column 4 we 
indicate which type of instanton branes were found. Five types are distinguished: $O1$, $O2$,  $S2$, $U1$ and $U2$, corresponding to $O(1)$, $O(2)$, $Sp(2)$, $U(1)$ and $U(2)$ Chan-Paton symmetry on the instanton volume. 
The number indicates the instanton brane multiplicity that gives the correct number of 
instanton charged zero modes from the {\bf a, b, c, d} branes, to lead to
 right-handed neutrino Majorana masses. The number of zero modes is in general
 the product of the instanton brane multiplicity and `intersection number' with the
 corresponding 4d spacefilling brane.  As discussed in Section \ref{themess},
 for symplectic branes the smallest possible brane multiplicity is 2.
As we discussed there, only $O1$ instantons may have the required universal minimal set of two zero modes in the uncharged sector. Still we look for all $O(1)$, $Sp(2)$ and $U(1)$ instantons which may yield a superpotential if the extra uncharged fermion
zero modes. In this vein we also include a search for $O2$ and $U2$ instantons.
 Note also that  such $O2$ or $U2$ instantons imply the existence of other
 instantons involving the same boundary state, but with multiplicity 1,
 which may lead to the R-parity violating operator $LH$. We will discuss the generation
of R-parity violating operators at the end of this section. The third character ($+$ or $-$) in the instanton in table \ref{tbl:TableInstanton} is the sign of $I_{Mc'} - I_{Mc}$. 
 For the instantons giving rise to right-handed neutrino Majorana masses
 this sign should be negative, whereas it should be positive for instantons giving rise to the Weinberg operator (or the $LH$ operator), see table \ref{tableoperators}.

The 1315 instantons are divided in the following way over the different types:
3 of types $O1+$ and $O1-$, 46 of type $U1+$, 24 of type $U1-$, 550 $S2+$, 627 $S2-$,
27 of types $U2+$ and $U2-$ and four of types $O2+$ and $O2-$. 
Notice that the vast majority (97.5\%) of the instanton solutions are of type $S2+$ and $S2-$.
 This is encouraging given the nice properties of such instantons, concerning e.g. R-parity conservation.
Note also that in almost all cases both $S-$ and $S+$ are simultaneously 
present\rlap,\footnote{In some models contributing many instantons there is
an {\it exact} symmetry between $S-$ and $S+$. This explains the {\it approximate} symmetry 
in the full set.  In some cases this symmetry can be
understood in terms of flipping the degeneracy labels of boundary states. We regard it
as accidental, since it is not found in all models.}
so both
sources of physical neutrino Majorana masses (from the see-saw mechanism or the Weinberg operator)
are present. 
The other instanton classes possibly  generating right-handed neutrino masses
are $O1-$ and $U1-$, which are much less abundant.
 There is just one orientifold with $O1-$ instantons,  for which one can obtain cancellation of RR tadpoles, see 
below. On the other hand we have found no orientifold with $U1-$ instantons and
cancellation of tadpoles, see below.

\LTcapwidth=14truecm
\begin{center}
\begin{longtable}{|l|l|l|l|l|l|}\caption{\em Summary of instanton branes.}\label{tbl:TableInstanton}\\
 \hline \multicolumn{1}{|l|}{Tensor}
& \multicolumn{1}{l|}{MIPF}
& \multicolumn{1}{l|}{Orientifold}
& \multicolumn{1}{l|}{Instanton}
& \multicolumn{1}{l|}{Solution}
\\ \hline
\endfirsthead
\multicolumn{5}{c}%
{{\bfseries \tablename\ \thetable{} {\rm-- continued from previous page}}} \\
\hline  \multicolumn{1}{|l|}{Tensor}
& \multicolumn{1}{l|}{MIPF}
& \multicolumn{1}{l|}{Orientifold}
& \multicolumn{1}{l|}{Instanton}
& \multicolumn{1}{l|}{Solution}
\\ \hline
\endhead
\hline \multicolumn{5}{|r|}{{Continued on next page}} \\ \hline
\endfoot
\hline \hline
\endlastfoot
(1,16,16,16) & 12 & 0          &   $S2^+$, $S2^-$        &   Yes           \\ \hline
(2,4,12,82) &  19 &     0       &   $S2^-!$               &    ?    \\
(2,4,12,82) &  19 &     0       &   $U2^+!, U2^-!$               &    No    \\
(2,4,12,82) &  19 &     0       &   $U1^+, U1^-$               &    No   \\ \hline
(2,4,14,46)  & 10 & 0          &           &              \\
(2,4,14,46)  &  16 &      0      &           &              \\ \hline
(2,4,16,34)  & 15  &      0      &              &               \\
(2,4,16,34)  & 15  &      1      &           &              \\
(2,4,16,34)  & 16  &      0      &  $S2^+$, $S2^-$         &       Yes       \\
(2,4,16,34)  & 16  &      1      &           &              \\
(2,4,16,34) & 18 & 0     &    $S2^-$               &    Yes          \\
(2,4,16,34) & 18 & 0     &    $U1^+, U1^-, U2^+, U2^-$      &   No          \\
(2,4,16,34)  & 49   &   0         & $U2^+, S2^-!, U1^+$          &      Yes        \\
(2,4,16,34)  & 49   &   0         & $U1^-$          &      No        \\ \hline
(2,4,18,28) & 17    & 0  &                    &              \\ \hline
(2,4,22,22) & 13 & 3            &  $S2^+!$, $S2^-!$          &   Yes!           \\
(2,4,22,22) & 13 & 2            &   $S2^+!$, $S2^-!$         &   Yes           \\
(2,4,22,22) & 13 & 1            &  $S2^+$, $S2^-$         &     No         \\
(2,4,22,22) & 13 & 0            &   $S2^+$, $S2^-$         &    Yes         \\
(2,4,22,22) & 31  &    1        & $U1^+, U1^-$      &      No        \\
(2,4,22,22) & 20  &     0       &          &         \\
(2,4,22,22)  & 46  &   0         &           &              \\
(2,4,22,22) & 49  &    1        &  $O2^+, O2^-, O1^+, O1^-$          &       Yes       \\ \hline
(2,6,14,14)  & 1  & 1          &    $U1^+$       &   No         \\
(2,6,14,14) & 22  &    2        &           &              \\
(2,6,14,14)   & 60  &    2        &           &              \\
(2,6,14,14)   & 64  &    0        &           &              \\
(2,6,14,14) & 65  & 0                 &           &     \\ \hline
(2,6,10,22) & 22  &      2      &           &              \\ \hline
(2,6,8,38)  & 16  &     0       &           &              \\ \hline
(2,8,8,18)  &  14  &     2       &    $S2^+!$, $S2^-!$         &   Yes           \\
(2,8,8,18)  &  14  &     0       &    $S2^+!$, $S2^-!$        &    No          \\ \hline
(2,10,10,10) & 52  &    0        &  $U1^+, U1^-$         &      No        \\ \hline
(4,6,6,10)  & 41  &    0        &           &              \\ \hline
(4,4,6,22)  & 43  &    0        &           &              \\ \hline
(6,6,6,6)    & 18  &   0         &           &              \\
\end{longtable}
\end{center}

Most models have a hidden sector containing extra boundary states beyond the SM ones.
In the same spirit of imposing chiral conditions first, we should require that $I_{Mh}=I_{Mh'}$,
 where $h$ is a hidden sector brane. This is to guarantee that the generated superpotential
does not violate some hidden sector gauge symmetry which would require the presence of
hidden sector fields along with the $\nu_R$ bilinear.
The latter condition is not imposed on the previously 
 known hidden sector ({\it i.e.} the one in \cite{schell1,schell2}), but instead  a new search for tadpole solutions was performed, for each $M$, restricting the candidate hidden sector branes to those satisfying $I_{Mh}=I_{Mh'}$ (as discussed in Section \ref{themess}).
This is because in general the known hidden sector in \cite{schell1,schell2}  
is just a  sample out of a huge number of possibilities.

In column 5 we indicate for which instantons it was possible to 
satisfy the tadpole conditions with this additional constraint. 
With regard to observable-hidden matter we
use the same condition as in \cite{schell1},
 namely that it is allowed only if it is
vector-like. Such a solution could be found for 879 of the 1315 instantons, with ten cases
inconclusive ({\it i.e} it was computationally too difficult to decide if a solution does or does not exist).
The latter are indicated with a question mark in column 5 (for most of the undecidable cases there is
a tadpole solution for a different instanton with the same characteristics; for that reason just
one question mark appears).
 
Independently of the RR tadpole condition (since there may be alternative sources for
 its cancellation, or hidden sectors which fall beyond the reach of RCFT), we can also 
consider the further constraint that the number of charged fermion zero modes is exactly
 right, not just in the chiral sense. This means $I_{M{\bf a}}=I_{M{\bf a}'}=I_{M{\bf b}}=I_{M{\bf b}'}=0$,
$I_{M{\bf c}}=2, I_{M{\bf c}'}=0$ and $I_{M{\bf d}}=-2, I_{M{\bf d}'}=0$ or vice-versa. Furthermore we
require that there are no adjoint or rank-2 tensor zero-modes (note that the latter could be chiral
if the instanton brane is complex, and indeed they are in some of the 1315 cases).
This reduces the 1315 instantons to 263. In column 4 we indicate those cases with an exclamation mark.
It is noteworthy that the success rate for solving the tadpole conditions is highest for these
instantons: 254 of the 263 allow a solution (with 3 undecided).
If an exclamation mark appears in column 4, this only indicates that {\it some} of the instantons are free of
the aforementioned zero modes, not that all of them are. But in all cases, if there are tadpole solutions, they
exist in particular for the configurations with an exclamation mark.
Finally
we may impose the condition that $I_{Mh}$ and $I_{Mh'}$ are separately zero. This is indicated with
and exclamation mark in column 5. This turns out to be very restrictive.
The only cases where this happens have no hidden sector
at all.

It is worth remarking that the only instantons having exactly the correct set of charged
zero modes and cancelling tadpoles are of $S2^{\pm}$ type. Also those instantons 
are  the only cases marked with an exclamation mark in column 4 and 5. 
These examples, which will be discussed below in some detail,
 also have just the minimal set  of fermion zero modes, except for the
 universal sector (which for $Sp(2)$ instantons contains two extra triplets).

The main conclusion about this scan  is that we did not find any instantons
 with exactly the zero mode fermions to generate the neutrino mass superpotential.
However we have found a number of examples which come very close to that, 
with  exactly the required charged zero modes and 
a very reduced set of extra uncharged zero modes from the universal sector.
These extra zero modes are non-chiral and hence one expects that e.g.
RR/NS fluxes or other effects may easily lift them, as we discussed in section 2.
Concerning $O(1)$ instantons, which have just the 
two required fermion zero modes in the universal sector,
 we have found one example, with the appropriate net structure
 of charged zero modes. However, it has plenty of other extra zero modes. 
We discuss examples of  $O(1)$ and $Sp(2)$ instantons in the following subsections.

\subsection{An O1 example}

Let us first discuss the case of $O(1)$ instantons.
In principle they would be the more attractive  since they have
no undesirable universal zero modes at all. Unfortunately this type of instanton
is rare within the set  we scanned, and we found just one example with a solution
to the tadpole equations without any unwanted chiral zero-modes. The instanton however has a very large number of uncharged and charged vector-like zero modes. 

The standard model brane configuration occurs for tensor product
 $(2,4,22,22)$, MIPF 49, orientifold 1, boundaries {\bf (a,b,c,d)} $ = (487,1365, 576,486)$. As usual we only provide this information in order to locate this model in the
 database. Further details are available on request.

 The bi-fundamental fermion spectrum of this model in the {\bf (a,b,c,d)} sector
 is fairly close to the MSSM:  there is an extra up-quark mirror pair, two mirror pairs
 of lepto-quarks with down quark charges and one with up-quark charges, plus two
 extra right-handed  neutrinos ({\it i.e.} a total of five right-handed neutrinos). There are
 three MSSM Higgs pairs.
The tensor spectrum is far less appealing, in particular for brane {\bf c}: this
 has 25 adjoints and 7 vector-like pairs of anti-symmetric tensors.

 As we said,  there is just one instanton brane of type $O1-$. It has exactly the
 right number of zero-modes with brane {\it d}, but five superfluous pairs of
 vector-like zero-modes with brane {\it c}, plus one vector-like pair with brane {\it a}.
 In addition there are four symmetric tensor zero-modes on the
 instanton brane (which of course are vector-like, since it is a real brane): the
 parameter $p_{+}$ in table 3 is equal to 2.

 The tadpole solution that is (chirally speaking) compatible with this instanton
 has a large hidden sector:  $O(1) \times O(2)^4 \times O(3) \times U(1)^2 \times Sp(2)^2 \times U(3)$
 (there are other possibilities, but no simple ones).  This hidden sector
 introduces more undesirable features: vector-like observable/hidden matter,
 vector-like instanton/hidden sector modes, plus chiral and non-chiral matter within
 the hidden sector.
Finally the coupling ratios are as follows: $\alpha_3/\alpha_2=.54$, ${\rm sin}^2 \theta_w=.094$,
 and the instanton coupling is 3.4 times weaker than the QCD coupling
 ($\alpha_{3}/\alpha_{\rm Instanton}= 3.4$).

Despite these unappealing features this model does demonstrate the existence
of this kind of solution.

\subsection{The S2 models}

As we have mentioned, these are the examples which come closer to the minimal set of fermion zero modes.
As we see in Table \ref{tbl:TableInstanton}, all such instantons satisfying the criteria on the zero mode structure (except for the extra universal zero modes) appear for models based on the same CFT orientifold. It is the one obtained from the $(2,4,22,22)$ Gepner model with MIPF 13 and orientifold 3 in the table. The model is obtained as follows. 

\medskip

\subsubsection{The closed string sector}

We start with the tensor product (2,4,22,22).
This yields a CFT with 12060 primary fields, 48 of which are simple currents,
forming a discrete group ${\cal G}={\bf Z}_{12} \times {\bf Z}_{2} \times {\bf Z}_{2}$.
After taking into account the permutation symmetry of the last two factors, we find that this
tensor product has 54 symmetric MIPFs, and we choose one of them to build the model of interest.
For convenience we specify all quantities in terms of a standard minimal model notation, but
also in terms of the labelling of the computer program ``kac" that generates the spectrum. This
particular MIPF is nr. 13.
To build
it we choose a subgroup of ${\cal G}$, which is isomorphic to
${\cal H}={\bf Z}_{12} \times {\bf Z}_{2}$.
The generator of the ${\bf Z}_{12}$ factor is primary field nr. 1, $(0,0,0,\{24,-24,0\},\{24,20,0\})$, and
the ${\bf Z}_{2}$ factor is generated by primary field nr. 24,
$(0,0,0,0,\{24,20,2\})$. The representations are specified on a basis $({\rm NSR},k=2,k=4,k=22,k=22)$,
{\it i.e.} the boundary conditions of the NSR-fermions and the four minimal models in the
tensor product.
Here $0$ indicates the CFT vacuum, and for all other states we use the familiar $(l,q,s)$ notation
for the $N=2$ minimal models.
The first generator has conformal weight
$h=\frac{11}{12}$
and
has ground state dimension 1. The second has weight $h=\frac{11}{2}$ and has ground state
dimension 2: the ground state contains both $(0,0,0,0,\{24,20,2\})$ and
$(0,0,0,\{24,20,2\},0)$. The matrix $X$ defining the MIPF according to the prescription given in 
\cite{Gato-Rivera:1991ru}\cite{Gato-Rivera:1990za}\cite{Kreuzer:1993tf}
 is
\beqa
\pmatrix{\frac{1}{12} & 0 \cr 0 & \frac{1}{2}} \cr
\eeqa
This simple current modification is applied to the charge conjugation invariant of the tensor product.
This defines a MIPF that corresponds to an automorphism of the fusion rules, and that pairs all
the primaries in the CFT off-diagonally. The number of Ishibashi states, and hence the number of boundary
states is 1080. The MIPF is invariant under exchange of the two $k=22$ factors: this maps current 24 to itself,
and current 1 to current 11, which is also in ${\cal H}$. Hence this symmetry of the tensor product maps
${\cal H}$ into itself, and it also preserves the matrix $X$.

To define an orientifold, we must specify a ``Klein bottle current"
plus two signs defined on the basis of the simple current group. For the current $K$ we use
the generator of the second ${\bf Z}_2$ in ${\cal G}$, primary field nr. 12.  This is the representation
$(0,0,\{4,-4,0\},\{(24,16,2)\},\{(24,-12,2)\})\, $ which is degenerate with nine other
states, all of dimension 1 and conformal weight 7. The crosscap signs are chosen, on the aforementioned
basis of ${\cal H}$ as $(+,-)$. This results in a crosscap coefficient of $0.0464731$, and it is orientifold nr. 3
of a total of 8. The orientifold is also invariant under permutation of the identical factors.

The closed string spectrum contains 14 vector multiplets and 60 chiral multiplets.

\medskip

\subsubsection{The standard model branes\label{SMbranes}}

To build a standard model configuration we have to specify the boundary state labels. It turns out that we
have four choices for label {\bf a} and {\bf b}, one for  {\bf c} and two for {\bf d}. This leads to a total
of 32 possibilities. Among these 32 there are 22 have distinct spectra (distinguished by the number of vector-like
states), but for all 32 choices one obtains the same set of dilaton couplings. It seems plausible
that these choices simply correspond to putting the {\bf a}, {\bf b} and  {\bf d} branes in slightly
different positions, so that we move the configuration in brane moduli space.
The choices are as follows (these are boundary labels assigned by the computer program, 
and can be decomposed in terms of minimal model representations; this will be explained
in table \ref{tbl:SMBoundaries} below)
\begin{eqnarray*}
{\bf a}:  &10,22,130,142  \\
{\bf b}:  &210,282,290,291  \\
{\bf c}:  &629  \\
{\bf d}:  &712,797  
\end{eqnarray*}
There are additional possibilities, but they do not give rise to
additional distinct spectra.

\vskip .3truecm
\begin{table}[h]\caption{\em Branes appearing in standard model configurations}
\label{tbl:SMBoundaries}
\begin{center}
~~~~~~~~~~~~~~~~\begin{tabular}{|l|c|c|c|c|c|} \hline
Label & Orbit/Deg. &  Reps & Weight & Dimension   \\ \hline \hline
 10  & 240 & $(0,0,0,0,\{10,0,0\})$ &   5/4 & 1 \\
 130  & 2760 & $(0,0,0,\{10,0,0\},0)$ &   5/4 & 1\\
 22  & [528,0] & $(0,0,0,\{1,-1,0\},\{11,1,0\})$ &   3/2 & 1\\
    &    & $(0,0,0,\{1,1,0\},\{11,-1,0\})$ &   3/2 & 1\\
 142  & [3048,0] & $(0,0,0,\{11,-1,0\},\{1,1,0\})$ &   3/2 & 1\\
      &   & $(0,0,0,\{11,1,0\},\{1,-1,0\})$ &   3/2 & 1\\
 210  & 4248 & $(0,0,\{3,3,0\},\{3,-3,0\},\{9,-9,0\})$ &   1/2 & 1\\
 282  & 5760 &  $ (0,0,\{3,3,0\}\{9,-9,0\}\{3,-3,0\})$ &   1/2 & 1\\
 290  & [5952,0] & $(0,0,\{1,1,0\}\{9,7,0\}\{11,-11,0\})$ &   5/6 & 1\\
 291  & [5952,24] &  $(0,0,\{1,1,0\}\{9,7,0\}\{11,-11,0\})$ &   5/6 & 1\\
 629  & [9348,30] & $(0,(1,-1,0),0,\{9,9,0\}\{5,-3,0\}$ &   7/12 & 1 \\
 712  & [9852,0] & $(0,\{1,1,0\}\{3,-3,0\}\{1,1,0\}\{5,5,0\})$ &   1/2 & 2\\
   &   & $(0,\{1,1,0\}\{1,-1,0\}\{1,1,0\}\{5,-3,0\})$ &   1/2 & 2\\
 797  & [10356,30] & $(0,\{1,1,0\}\{3,-3,0\}\{5,5,0\}\{1,1,0\})$ &   1/2 & 2 \\
    &   & $(0,\{1,1,0\}\{1,-1,0\}\{5,-3,0\}\{1,1,0\})$ &   1/2 & 2 \\ \hline 
\end{tabular}
\end{center}
\end{table}

The second column gives the boundary labels in terms of a primary field label and a degeneracy
label (boundaries not indicated by square brackets are not degenerate).
The labels appearing in columns 1 and 2 are assigned by the computer program,
and are listed here only for the purpose of reproducing the results using that program.
 In column
2, the boundary labels are expressed in terms of primary field labels, as in
formula (\ref{Blbl}).
If a single number appears, this is a representative of an ${\cal H}$-orbit corresponding
to the boundary.
If square
brackets are used, this means that the ${\cal H}$-orbit has fixed points, and that
it corresponds to more than one boundary label. The second entry in the square
brackets is the degeneracy label, and refers to a character of the ``Central Stabilizer"
defined in \cite{Fuchs:2000cm}; the details of the definition and the labelling will not
be important here. In this case the first entry within the square brackets refers to an
orbit representative.

These orbit representatives can also be expressed in a standard
form for minimal model tensor products.
This is done in column 3. This is basically the same expansion
shown in (\ref{Blbl}), except that the degeneracy label $\Psi_I$ turns
out to be trivial in all cases, both for the standard model and for the
instanton branes  shown below (although the theory does contain primaries with
non-trivial $\Psi$'s).
In columns 4 and 5 we specify the weight and ground state dimension
of the corresponding highest weight representation. These data are not directly
relevant for the boundary state, but helps in identifying it.

Since boundaries are specified by orbit representatives, it is not straightforward
to compare them, since the standard choice (the one listed in column 2) is arbitrary.
For this reason we have used another representative in columns
3, 4 and 5, selected by an objective criterion: we choose the one of minimal dimension
and minimal conformal weight (in that order). If there is more than one representative satisfying
these criteria we list all.

\subsubsection{The open string spectrum}

In Table \ref{tbl:TableSpectra} we summarize the spectra of the 32 models. The first four 
columns list the {\bf a,b,c,d} brane labels. The last eight columns specify the
total number of multiplets of types Q (quark doublet), U (up quark singlet), D
(down quark singlet), L (lepton doublet), E (charged lepton singlet), N (neutrino
singlet), Y (lepto-quark) and H (Higgs). The numbers given are for the total number
of lefthanded fermions in the representation, plus their complex conjugates. So for example
a 7 in column ``Q" means that there are 5 quark doublets in the usual representation
$(3,2,\frac16)$, plus two in the complex conjugate representation $(3^*,2,-\frac16)$.

This yields the required three families of quark doublets, plus two mirror pairs. Hence the
smallest number that can occur in the six columns QUDLEN is three, if there
are no mirrors (note that cubic anomaly cancellation requires three right-handed
neutrinos in this class of models). The lepto-quarks Y are all in the same representation
as the down-quarks (D), or the conjugate thereof, and they occur only as vector-like
mirror pairs. They differ from D-type mirror quarks because they carry lepton number, because
they come from open strings ending on the {\bf d}-brane instead of the {\bf c}-brane.
In general, there can also exist U-type lepto-quarks, but in these models they do
not occur. 
Finally the numbers 10, 18 and 26 in column 'H' mean that there are 
5, 9 or 13 MSSM Higgs pairs $H+{\bar H}$. It is worth noticing that right-handed quarks $U,D$ and
neutrinos $N=\nu_R$ do not have vectorlike copies. On the other hand right-handed leptons 
$E$ always have one  and the left-handed fields $Q,L$ may have up to 3
vector-like copies.

\LTcapwidth=14truecm
\begin{center}
\begin{longtable}{|c|c|c|c|c|c|c|c|c|c|c|c|}\caption{\em Spectrum all 32 configurations.}\label{tbl:TableSpectra}\\
 \hline
  \multicolumn{1}{|l|}{$U(3)$} 
& \multicolumn{1}{l|}{$Sp(2)$}
& \multicolumn{1}{l|}{$U(1)$}
& \multicolumn{1}{l|}{$U(1)$}
& \multicolumn{1}{l|}{Q}
& \multicolumn{1}{l|}{U}
& \multicolumn{1}{l|}{D}
& \multicolumn{1}{l|}{L}
& \multicolumn{1}{l|}{E}
& \multicolumn{1}{l|}{N}
& \multicolumn{1}{l|}{Y}
& \multicolumn{1}{l|}{H}
\\ \hline
\endfirsthead
\multicolumn{12}{c}%
{{\bfseries \tablename\ \thetable{} {\rm-- continued from previous page}}} \\
\hline   \multicolumn{1}{|l|}{$U(3)$}
& \multicolumn{1}{l|}{$Sp(2)$}
& \multicolumn{1}{l|}{$U(1)$}
& \multicolumn{1}{l|}{$U(1)$}
& \multicolumn{1}{l|}{Q}
& \multicolumn{1}{l|}{U}
& \multicolumn{1}{l|}{D}
& \multicolumn{1}{l|}{L}
& \multicolumn{1}{l|}{E}
& \multicolumn{1}{l|}{N}
& \multicolumn{1}{l|}{Y}
& \multicolumn{1}{l|}{H }
\\ \hline
\endhead
\hline \multicolumn{12}{|r|}{{Continued on next page}} \\ \hline
\endfoot
\hline \hline
\endlastfoot
 10 & 210 & 629 & 712 & 7 & 3 & 3 & 9  & 5 & 3 & 6 & 10 \\
 22 & 210 & 629 & 712 & 7 & 3 & 3 & 9 & 5 & 3 & 6 & 10\\
 130 & 210 & 629 & 712 & 3 & 3 & 3 & 9 & 5 & 3 & 2 & 10\\
 142 & 210 & 629 & 712 & 3 & 3 & 3 & 9 & 5 & 3 & 2 & 10\\
 10 & 282 & 629 & 712 & 3 & 3 & 3 & 5 & 5 & 3 & 6 & 26\\
 22 & 282 & 629 & 712 & 3 & 3 & 3 & 5 & 5 & 3 & 6 & 26\\
 130 & 282 & 629 & 712 & 7 & 3 & 3 & 5 & 5 & 3 & 2 & 26\\
 142 & 282 & 629 & 712 & 7 & 3 & 3 & 5 & 5 & 3 & 2 & 26\\
 10 & 290 & 629 & 712 & 3 & 3 & 3 & 3 & 5 & 3 & 6 & 18\\
 22 & 290 & 629 & 712 & 3 & 3 & 3 & 3 & 5 & 3 & 6 & 18\\
 130 & 290 & 629 & 712 & 3 & 3 & 3 & 3 & 5 & 3 & 2 & 18\\
 142 & 290 & 629 & 712 & 3 & 3 & 3 & 3 & 5 & 3 & 2 & 18\\
 10 & 291 & 629 & 712 & 3 & 3 & 3 & 3 & 5 & 3 & 6 & 18\\
 22 & 291 & 629 & 712 & 5 & 3 & 3 & 3 & 5 & 3 & 6 & 18\\
 130 & 291 & 629 & 712 & 3 & 3 & 3 & 3 & 5 & 3 & 2 & 18\\
 142 & 291 & 629 & 712 & 3 & 3 & 3 & 3 & 5 & 3 & 2 & 18\\
 10 & 210 & 629 & 797 & 7 & 3 & 3 & 5 & 5 & 3 & 2 & 10\\
 22 & 210 & 629 & 797 & 7 & 3 & 3 & 5 & 5 & 3 & 2 & 10\\
 130 & 210 & 629 & 797 & 3 & 3 & 3 & 5 & 5 & 3 & 6 & 10\\
 142 & 210 & 629 & 797 & 3 & 3 & 3 & 5 & 5 & 3 & 6 & 10\\
 10 & 282 & 629 & 797 & 3 & 3 & 3 & 9 & 5 & 3 & 2 & 26\\
 22 & 282 & 629 & 797 & 3 & 3 & 3 & 9 & 5 & 3 & 2 & 26\\
 130 & 282 & 629 & 797 & 7 & 3 & 3 & 9 & 5 & 3 & 6 & 26\\
 142 & 282 & 629 & 797 & 7 & 3 & 3 & 9 & 5 & 3 & 6 & 26\\
 10 & 290 & 629 & 797 & 3 & 3 & 3 & 3 & 5 & 3 & 2 & 18\\
 22 & 290 & 629 & 797 & 3 & 3 & 3 & 3 & 5 & 3 & 2 & 18\\
 130 & 290 & 629 & 797 & 3 & 3 & 3 & 3 & 5 & 3 & 6 & 18\\
 142 & 290 & 629 & 797 & 3 & 3 & 3 & 3 & 5 & 3 & 6 & 18\\
 10 & 291 & 629 & 797 & 3 & 3 & 3 & 3 & 5 & 3 & 2 & 18\\
 22 & 291 & 629 & 797 & 5 & 3 & 3 & 3 & 5 & 3 & 2 & 18\\
 130 & 291 & 629 & 797 & 3 & 3 & 3 & 3 & 5 & 3 & 6 & 18\\
 142 & 291 & 629 & 797 & 3 & 3 & 3 & 3 & 5 & 3 & 6 & 18\\ 
\end{longtable}
\end{center}

In the following table we list the multiplicities $L_{aa}$ and
$L_{aa'}$ of the branes that occur in these models, leading to vector-like sets of adjoints and rank-2 tensors. Since brane {\bf b} is symplectic, the
number of adjoints is equal to the number of symmetric tensors.

\LTcapwidth=14truecm
\begin{center}
\begin{longtable}{|c|c|c|c|}\caption{\em 4d matter from the $aa$ and $aa'$ sectors.}\label{tbl:TableDiagonalSpectra}\\
 \hline
  \multicolumn{1}{|l|}{Boundary} 
& \multicolumn{1}{l|}{Adjoints}
& \multicolumn{1}{l|}{Anti-symm.}
& \multicolumn{1}{l|}{Symm.}
\\ \hline
\endfirsthead
\multicolumn{4}{c}%
{{\bfseries \tablename\ \thetable{} {\rm-- continued from previous page}}} \\
\hline 
  \multicolumn{1}{|l|}{Boundary} 
& \multicolumn{1}{l|}{Adjoints}
& \multicolumn{1}{l|}{Anti-symm.}
& \multicolumn{1}{l|}{Symm.}
\\ \hline
\endhead
\hline \multicolumn{4}{|r|}{{Continued on next page}} \\ \hline
\endfoot
\hline \hline
\endlastfoot
{\bf a}(10) & 2  &  2  &  6\\
{\bf a}(22) & 2  &  2  &  2\\
{\bf a}(130)  &  2  &  2  &  6\\
{\bf a}(142)  &  2  &  2  &  2\\
{\bf b}(210) & - & 14 & 10\\
{\bf b}(282) & -  & 14 & 10\\
{\bf b}(290) & - & 14 & 6\\
{\bf b}(291) &-  & 14 & 6\\
{\bf c}(629)  &  9  &  -  &  14\\
{\bf d}(712)  &  3  &  -  &  6\\
{\bf d}(797)  &  3  &  -  &  6\\
\end{longtable}
\end{center}

\medskip

It should be emphasized that CFT constructions generically correspond to
particular points in moduli space of CY orientifolds. Due to this, they 
usually have an `enhanced' massless particle content with extra vector-like 
matter and closed string gauge interactions. Thus one would expect that
many of the massless vector-like chiral fields present in this class of models 
could gain masses while moving to a nearby point in moduli space.

\subsubsection{The instantons}

Each of these 32 Standard Model compactifications admits 8 instantons. The instanton labels are identical for all the 32 models.
They are listed  in Table \ref{tbl:InstantonBranes}.
 The first five columns use the same notation
as for the standard model boundary labels.
In column 6
we list the numerical value of the dilaton coupling to the instanton brane. This quantity
is proportional to ${1\over g^2}$. It is instructive to compare 
these couplings to the gauge couplings, in order to gain intuition
 on the suppression factor for our instantons. In these models the $U(3)$ dilaton
 couplings are $0.00622$, so that the instantons are more strongly coupled
than QCD\footnote{ Note that the Type II dilaton in this
compactifications is an arbitrary parameter which can always be chosen so that 
we consistently work at weak coupling. It is the relative value of 
gauge couplings which we are comparing here.} On the other hand in this particular model 
the ratio $\alpha_3/\alpha_2$ at the string scale is 3.23 (the value of ${\rm sin^2} \theta_w$ at the
string scale is 0.527). All of these couplings are subject to renormalization group running, 
and there are plenty of vector-like states to contribute to this, if one assumes that they 
acquire masses at a sufficiently low scale. One should perform a detailed renormalization group analysis
to check whether one may obtain consistency with the gauge couplings measured at
low-energies. Let us emphasize however that one expects that moving in moduli space
many of these vector-like states will gain masses and also the values of the different 
gauge couplings will also generically vary. 

Since the value of the Type II dilaton is a free parameter at this level,
one can get the appropriate (intermediate) mass scale for the right-handed
neutrino Majorana masses by choosing an appropriate value for the dilaton. 
In this context,  it is satisfactory to verify that the instanton couplings are unrelated to 
the gauge couplings, as expected since they do not correspond to gauge instantons \cite{iu}, and are in fact less suppressed than the latter.

\vskip .3truecm
\renewcommand{\arraystretch}{1.2}
\begin{table}[h!]\caption{\em Instantons for all 32 configurations}\label{tbl:InstantonBranes}
\begin{center}
~~~~~~~~~~~~~~~~\begin{tabular}{|l|c|c|c|c|c|} \hline
Lbl. & Orbit/Deg. &  Reps & Weight & Dim. & coupling  \\ \hline \hline
414 & [8064,0]  &  $(0,\{1,1,0\},0,\{22,-22,0\},\{20,16,0\})$  & 5/2 & 1 & 0.0016993 \\
417 & [8076,30] &   $(0,\{1,-1,0\},0,\{22,22,0\},\{20,-16,0\})$ & 5/2 & 1 & 0.0016993 \\
456 & [8316,0]  &   $(0,\{1,1,0\},0,\{20,16,0\},\{22,-22,0\})$  & 5/2 & 1 & 0.0016993\\
459 & [8328,30] &   $(0,\{1,-1,0\},0,\{20,-16,0\},\{22,22,0\})$  & 5/2 & 1&  0.0016993\\
418 & [8088,0]  &    $(0,\{1,1,0\},0,\{22,-22,0\},\{18,16,0\})$  & 5/3 & 1 & 0.0027033\\
420 & [8100,0]  &   $(0,\{1,-1,0\},0,\{22,22,0\},\{18,-16,0\})$  & 5/3 & 1 & 0.0027033\\
502 & [8592,0]  &    $(0,\{1,1,0\},0,\{18,16,0\},\{22,-22,0\})$  & 5/3 & 1 & 0.0027033\\
505 & [8604,30] &    $(0,\{1,-1,0\},0,\{18,-16,0\},\{22,22,0\})$ & 5/3 & 1 & 0.0027033\\ \hline
\end{tabular}
\end{center}
\end{table}

\renewcommand{\arraystretch}{1.0}

Note that the 8 instantons  fall into two distinct classes (evidently not related by 
any discrete symmetry, since the
conformal
weight on the boundary orbit is distinct, and the coupling is different as well).
Within each class, the orbits of the four instanton boundaries
appear to be
 related by the ${\bf Z}_2$ symmetries of interchange of the last two tensor factors, and simultaneous
inversion of the charge $q$ of the minimal model. However, one has to be very careful
in reading off symmetries directly from the labels in columns 3 of Tables (\ref{tbl:SMBoundaries}) and 
(\ref{tbl:InstantonBranes}) for a number of reasons. First of all the entries in column 3 are
representatives of boundary orbits, and these representatives themselves are merely
representatives of extension orbits. Secondly the action of any discrete symmetry
on the degeneracy labels can be non-trivial. In appendix B we discuss these symmetries
in more detail.

\subsection{Other examples}
The $Sp(2)$ instanton examples just discussed are the ones which get closer
to the required minimal set of fermion zero modes.
Under slightly weaker conditions, we find many
more solutions. In all these cases some additional mechanism beyond exact RCFT will
be needed to lift some undesirable zero modes. 

The simplest such case is the 
following. The tensor product is $(2,8,8,18)$, MIPF nr. 14, orientifold 2 (the precise
spectra may be found using this information in the database  \href{http://www.nikhef.nl/~t58/filtersols.php}{www.nikhef.nl/$\sim$t58/filtersols.php}).
There are
three distinct brane configurations for which almost perfect instantons exist,
namely $({\bf a},{\bf b},{\bf c},{\bf d})=(64,562,389,67)$
and $(64,577,389,67)$ and $(65,560,189,66)$. Each has six instantons, three  of
type
S2+ and three of type S2$-$. As in the foregoing example, the six instantons are
identical for the three standard model configurations. In this example, they have
three different dilaton coupling strengths: $.00254, .00665$ and $.0108$ (each value
occurs once for $S2+$ and once for $S2-$). By comparison, the $U(3)$-brane dilaton
 coupling strength is 0.0119338, so that the instanton brane coupling is quite a bit
stronger than the QCD coupling. This is again an interesting  point if we want that 
$\nu_R$ masses are not too much suppressed. 
Furthermore in this example there are three distinct
instanton couplings, so that one may expect three non-zero eigenvalues (with a hierarchy) in the
mass matrix. As in the previous examples there is not gauge coupling unification, one rather has
$\alpha_3/\alpha_2=.4813$ and $\sin^2(\theta_w)=.183$ at the string scale.
Again a full renormalization group analysis should be performed in order to 
check consistency with the measured low-energy gauge coupling values. 

These models all have a hidden sector consisting of a single $Sp(2)$ factor. They have
respectively 3, 1 and 3 susy Higgs pairs, 
and a spectrum of bi-fundamentals that is  closer to
that of the standard model than the previously discussed
$Sp(2)$ examples: quarks and leptons do not have vector-like copies 
(there are only some vector-like leptoquarks), and even one of the three 
models have the minimal set of Higgs fields of the MSSM.
 The rest of the spectrum is purely vector-like, and
contains a number of  rank-2 tensors, including eight or six adjoints 
of $U(3)$. Furthermore there is vector-like observable-hidden matter. 
The only undesirable instanton zero-mode is a single bi-fundamental
 between the hidden sector $Sp(2)$ brane and the instanton brane.
 Still, these SM brane configurations without the hidden sector, provide interesting and very simple local models of D-brane sectors admitting instantons generating neutrino masses 
(with the additional ingredients required to eliminate the extra universal triplets of fermion zero modes).

\subsection{R-parity violation}

We now turn to the generation of other possible superpotentials 
violating $B-L$.
An instanton violates R-parity if the amount of $B-L$ violation,
\beqa
 I_{M{\bf a}}-I_{M{\bf a}'}-I_{M{\bf d}}+I_{M{\bf d}'} 
 \eeqa
 is odd. Examples of instantons with that property were found
 in the following tensor product/MIPF/orientifold combinations:
$[(1,16,16,16), 12,0]$, $[(2,4,16,34), 49, 0]$, $[(2,4,12,82),19,0]$
$[(2,4,22,22) ,49,0]$ and $[(2,4,16,34),18,0]$.  Note that all cases
for which $O2$ or $U2$ instantons were found necessarily have R-parity
violating instantons as well: the corresponding $O1$ and $U1$ instantons
have $I_{M{\bf d}}$ or $I_{M{\bf d}'}$ equal to $\pm 1$, whereas the intersection with
the {\bf a} is non-chiral. In principle, there are many more ways to obtain R-parity
violating instantons (either due to non-vanishing contributions to 
$I_{M{\bf a}}-I_{M{\bf a}'}$ or higher values of $I_{M{\bf d}}-I_{M{\bf d}'}$), and indeed,
many such instantons turn out to exist. But the number of 
tensor product/MIPF/orientifold combinations where they occur hardly increases:
only in the case $[(1,16,16,16), 12,0]$ we found R-parity violating instantons, but no
$U1$ or $O1$ instantons. This suggests that in the
other cases R-parity is a true symmetry of the model. Unfortunately we have no way
of rigorously ruling out any other non-perturbative effects, but at least the set we can
examine respects R-parity. This includes in particular the models without
hidden sector (found for $[(2,4,22,22),13,3]$ ) discussed above.

The following table list the total number of instantons with the
chiral intersections listed in table \ref{tableoperators}. The total number of instantons
(boundaries violating the sum rule, as defined in (\ref{AnomalyCond})) is 29680, for
all standard model configurations combined. The last four columns indicate how
many unitary instantons  satisfy the sum rule exactly as listed in table (\ref{AnomalyCond}), 
how many satisfy it with $I_{M{\bf x}} \leftrightarrow -I_{M{\bf x}'}$ (the column U'),  and how
many O-type and S-type instantons there are. Here `S' refers to boundaries with a
symplectic Chan-Paton group if the boundary is used as an instanton brane. All
intersection numbers for type S have been multiplied by 2 before comparing with
table \ref{tableoperators}. For real branes, the relevant quantities used in the
comparison
are $I_{M{\bf a}}-I_{M{\bf a}'}$, $I_{M{\bf c}}-I_{M{\bf c}'}$ and $I_{M{\bf d}}-I_{Md'}$, while $I_{M{\bf b}}=0$.
There are fewer unitary instantons possibly generating Majorana masses then the numbers
mentioned above because the conditions we use here are stricter: we require here
that $I_{M{\bf x}}$ and $I_{M{\bf x}'}$ match exactly, not just their difference. Note however that
this still allows additional vector-like zero-modes. If we only wish to consider cases without {\it any}
spurious zero-modes, we may limit ourselves to the O-type instantons in the last column.
There are very few to inspect, and all of them turn out to have a few non-universal zero modes.

\vskip .7truecm
\begin{table}[htb]
\renewcommand{\arraystretch}{0.75}
\begin{center}
\begin{tabular}{|c||c|c|c|c|}
\hline $D=4$ Operator & U  &  U' &  S  &  O  \\
\hline\hline   $\nu_R\nu_R$ &  1  &  2   &  627   & 3  \\
\hline $L{\bar H}L{\bar H}$  & 0   &  5   &  550    & 3 \\
\hline\hline  $L{\bar H}$  &  3    &  3  &  0    & 4  \\
\hline $QDL$  &    8    &  4   &  0     & 4 \\
\hline $UDD$  & 0   &  0   &  0   &  4 \\
\hline $LLE$  &  8    &  4   &  0  & 4  \\
\hline \hline
 $QQQL$ &   0    &  4   &  0      & 3892  \\
\hline $UUDE$   &  4    &  0   &   0  & 3880 \\
\hline
\end{tabular}
\end{center}
\caption{\small Number of instantons in our search which may 
induce neutrino masses (first 2 rows), R-parity violation (next 4 rows)
or proton decay operators (last 2 rows).}
\label{tableoperatorsCount}
\end{table}

The last two cases are $B-L$ preserving dimension five operators, and obviously
do not come from the set of 29680 $B-L$ violating instantons. They were searched
for separately, but the search was limited to the same 391 models we used in the
rest of the paper.
Obviously, one could equally well look for such instantons in the full database, since
their existence does not  require a massive $B-L$.

 It is interesting to note that in the classes of MSSM-like models
discussed earlier in this section with the closest to minimal zero mode 
structure, there are no instantons al all  generating either R-parity violating 
or the $B-L$ dim=5 operators in the table. This makes them particularly attractive.

Note that all numbers in table \ref{tableoperatorsCount} refer to the occurrence of instantons 
in the set of 391 tadpole-free models with massive B-L, but without checking the presence
of zero-modes between the hidden sector and the instanton. It makes little sense to use
the hidden sector in the database for such a check, since this is just one sample from
a (usually) large number of possibilities. A meaningful question would be: can one find
a hidden sector that has no zero-modes with the instanton. We have done such a search for the
$B-L$ violating instantons (see the exclamation marks in the
last column of table (\ref{tbl:TableInstanton})), but not for the $B-L$ preserving instantons.

\section{Conclusions and outlook}

In this paper we have presented a systematic search for MSSM-like Type II Gepner orientifold models allowing for boundary states associated to instantons giving rise to neutrino Majorana masses. This search is very well  motivated since 
neutrino masses are not easily accommodated in the  semi-realistic compactifications 
constructed up to now.  String instanton induced Majorana masses provides a novel
and promising way to understand the origin of neutrino masses in the 
string theory context.

The string instantons under discussion are not gauge instantons. Thus, for example, 
they not only break $B+L$ symmetry (like 't Hooft instantons do) but also $B-L$,
allowing for Majorana neutrino mass generation. The obtained mass terms are  
of order $M_s \, \exp(-V/g^2)$ but this suppression is unrelated to the 
exponential suppression of e.g. electroweak instantons and may be mild. In fact we
find in our most interesting examples that the instanton action is typically substantially
smaller than that of QCD or electroweak instantons, and hence these effects are much less 
suppressed than those coming from gauge theory instantons.

To perform our instanton  search  we have  analyzed the structure of the zero modes  that these instantons must have  in order to induce the required superpotential. 
This analysis goes beyond the particular context of Gepner orientifolds and has general
validity for Type II CY orientifolds.
We have found that instantons with
$O(1)$ CP symmetry have the required universal sector of just two fermionic zero modes
for the superpotential to be generated. Instantons with $Sp(2)$ and $U(1)$ CP symmetries
have extra unwanted universal fermionic zero modes, which however may be
lifted in a variety of ways in more general setups, as we discuss in the text. 
In fact we find in our search that  around 98 \% of the instantons with the correct
structure of charged zero modes have  $Sp(2)$ CP symmetry.     
Indeed, from a number of viewpoints the $Sp(2)$ instantons are specially
interesting. The instantons we find with the simplest structure of 
fermionic zero modes are $Sp(2)$ instantons which are also the ones
which are present more frequently in the MSSM-like class of Gepner
constructions considered. They have also some interesting features from
the phenomenological point of view. Indeed, due to the non-Abelian structure of
the CP symmetry, the structure in flavor space of the neutrino Majorana masses
factorizes. This makes that, irrespective of what particular compactification is
considered, $Sp(2)$ instantons may easily lead to a hierarchical structure of
neutrino masses.
It would be important to further study the possible phenomenological 
applications of the present neutrino mass generating mechanism.

String instanton effects can also give rise to other B- or L-violating
operators. Of particular interest is the dimension 5 Weinberg operator giving
direct Majorana masses to the left-handed neutrinos. We find that in the 
most interesting cases, different instantons giving rise to the Weinberg operator 
and to $\nu_R$ Majorana masses are both simultaneously present.
Which effect is the dominant one in the generation of the physical
light neutrino masses depends on the values of the instanton actions 
and amplitudes as well as on the value of the string scale.
Instantons may also generate dim$<5$ operators violating  R-parity.
We find however that instantons inducing such operators are extremely rare, 
and in fact are completely absent in the  Gepner models 
with the simplest $Sp(2)$ instantons inducing neutrino masses.

There are many avenues yet to be explored. It would be important to understand
better the possible sources  (moving in moduli space, addition of RR/NS backgrounds etc.) of uplifting for the  extra uncharged fermionic zero modes  in the most favoured $Sp(2)$ instantons. A second important question is that we have concentrated on checking the existence of instanton zero modes appropriate to generate neutrino masses; one should
further check that the required couplings among the fermionic zero modes and
the relevant 4d superfields (i.e. $\nu_R$ or $L{\bar H}$) are indeed present
in each particular case. This is in principle possible in models with a known CFT
description but could be difficult in practice for the Gepner models here described.
 
Instantons can also generate other superpotentials with interesting  physical 
applications. One important example is the generation of a  Higgs 
bilinear  (i.e. a $\mu $-term)  in MSSM-like models \cite{bcw,iu}.
Thus, e.g., one could perform a systematic search for instantons 
(boundary states) generating
a $\mu $-term in the class of  CFT Gepner orientifolds considered in the 
present article. Other possible application is the search for instantons 
inducing superpotential couplings involving only closed string moduli. 
The latter may be useful for the moduli-fixing problem, or for non-perturbative corrections to perturbatively allowed couplings \cite{Abel:2006yk}. 

Finally, it would  be important to search for  analogous instanton effects inducing neutrino masses in  other string constructions (heterotic, M-theory etc.). A necessary condition is that
 the anomaly free $U(1)_{B-L}$ gauge boson  should become massive due to a St\"uckelberg term. 

The importance of neutrino masses in physics beyond the Standard Model is unquestionable. We have shown that string theory instantons provide an elegant and simple mechanism to implement them in semi-realistic MSSM-like string vacua, and a powerful constraint in model building. In our opinion, the conditions of the existence of appropriate instantons to generate neutrino masses should  be an important guide in a
search for a string description of the Standard Model.

{\bf Acknowledgements}\\
We thank M. Bertolini, R. Blumenhagen, S. Franco, M. Frau, S. Kachru, E. Kiritsis, A. Lerda, 
D. L\"ust, F. Marchesano, T. Weigand for useful discussions. A.M.U. thanks M. Gonz\'alez for encouragement and support.
The research of A.N. Schellekens was funded in part by program
FP 57 of the Foundation for Fundamental Research of Matter (FOM), and
Research Project FPA2005-05046 of de Ministerio de Educacion y Ciencia, Spain.
The research by L.E. Ib\'a\~nez and A.M. Uranga has been supported by the European 
Commission under RTN European Programs MRTN-CT-2004-503369,
 MRTN-CT-2004-005105, by the CICYT (Spain), and the Comunidad de Madrid under project HEPHACOS P-ESP-00346.

\newpage

\appendix

\vskip 10mm
 \renewcommand{\theequation}{\thesection.\arabic{equation}}
\centerline{\Large\bf Appendix}
\addcontentsline{toc}{section}{Appendix}

\section{CFT Notation}

Here we summarize the labelling conventions for various CFT quantities.
Further details and explanations can be found in \cite{Fuchs:2000cm}.

It is important to keep in mind that there are four steps in the construction,
each involving choices of some quantities. The steps are

\begin{itemize}
\item{A CFT tensor product}
\item{An extension of the chiral algebra of this tensor product}
\item{The choice of a MIPF}
\item{The choice of an orientifold}
\end{itemize}

The second and third step are easily confused. A MIPF can itself be of
extension type (although it can also be of automorphism or mixed type), 
meaning that it implies an extension of the chiral algebra. The crucial
difference between step two and three in that case is that in step 2 all 
fields that are non-local with respect to the extension are projected out, and
the symmetry of the extension is imposed on all states of the CFT, 
{\it i.e} in particular on all boundary states.  The extension
in step three acts as a bulk invariant, but  the boundary states are not required to 
respect the symmetry implied by the extension.

Primary fields  of $N=2$ minimal models are labeled in the usual
way by three integers $(l,q,s)$. In addition to these minimal models, one
building block of our CFT's is of course a set of NSR fermions in four
dimensions. They can be represented by the four conjugacy classes $(0),(v),(s),(c)$
analogous to those of a root lattice of type D. 

Primary fields in a tensor product of $M$ factors are therefore labelled
as
\beqa
\label{TensorChar}
I = ((x),(l_1,q_1,s_1),\ldots,(l_M,q_M,s_M))
\eeqa
where $x=0,v,s$ or $c$.

This tensor product is extended by the alignment currents and the spin-1 field
corresponding to the space-time supersymmetry generator. This organizes the tensor
product fields into orbits, which can be labelled by one of the elements of the orbit.
We always choose the field of minimal conformal weight (or one of them, in case
there are more) as the orbit representative labelling the orbit. The supersymmetry 
generator may have fixed points, leading to orbits appearing more than once
as primary fields of the extended theory. In those cases we need an additional
degeneracy label to distinguish them. It is convenient to choose for this label
a character of the discrete group that is causing the degeneracy, the ``untwisted
stabilizer", which depends on $I$.
Denoting this character as $\Psi_I$ we get then the following
set of  labels for the primaries of the extended CFT
\beqa
i=[I,\Psi_I] \,
\eeqa
where $I$ has the form (\ref{TensorChar}).
If there are no degeneracies we will leave out the square brackets and the $\Psi_I$.

In boundary CFT's two new labels appear: the labels of Ishibashi-states that
propagate in the transverse channel of the annulus, and the boundary labels. In
the simplest, ``Cardy" case both sets of labels are in one-to-one correspondence
with the extended CFT labels $i$. But if we consider non-trivial MIPFs $Z_{ij}$ both sets
of labels are different. The Ishibashi states are in one-to-one correspondence
with the fields $i$ with $Z_{ii^c} \not =0$. Degeneracies can occur here if 
$Z_{ii^c} \not > 0$. This requires the introduction of a degeneracy label.
Such degeneracies may occur if the stabilizer of $i$ (the set of simple currents
that fix $i$) is non-trivial. It is convenient to use elements $J$ of the stabilizer as
degeneracy labels, so that the Ishibashi labels get the following form
\beqa
m = (i,J) \ ,
\eeqa
where $i$ is an extended CFT label, as defined above (to be precise, in some cases 
a non-trivial degeneracy label is introduced even if $Z_{ii^c}=1$. The details will not matter here).

Boundary states correspond to orbits of the simple current group ${\cal H}$ that defines
a MIPF. To label such orbits we choose a representative. There is no obvious  
canonical representative (one could use one of minimal conformal weight, but
the conformal weight of orbit members of a boundary state does not play any r\^ole in
the formalism, unlike the conformal weight of a primary). So in this case we just make
an arbitrary choice. Once again there can be degeneracies. In this case they are due
to a subgroup of the stabilizer called the ``Central Stabilizer". It is convenient
to label the boundary states by an orbit representative $i$  and a character $\psi_i$
of the central stabilizer. If we expand the  boundary state label in all of its
components we get
\beqa
\label{Blbl}
a = [ i, \psi_i] = [[I,\Psi_I],\psi_i] =[ ((x),(l_1,q_1,s_1),\ldots,(l_M,q_M,s_M)),\Psi_I],\psi_i] \ .
\eeqa
Note that $i$ is just a  representative of a boundary orbit, and that $I$ is just
a representative of an orbit of the extension of the CFT.

\section{Instanton boundary symmetries}

In the hidden-sector free example discussed in some detail in 
section \ref{scan} we have
encountered  $Sp(2)$-type instantons, the most common kind in our scan.
This particular model is the one that comes closest to the required zero mode count,
although the only superfluous zero modes are rather awkward.
Let us assume that the effect of these superfluous universal zero-modes
instantons can be avoided. Then there is still another problem we have to face, namely
that the two zero modes $\alpha^i$ and
$\gamma^i$ are related by an $Sp(2)$ transformation of the label $i$.
Then we we need at least three
independent instantons (with unrelated couplings) to generate three non-zero  neutrino 
masses, as discussed in Section \ref{flavour}. Since the technology to compute
the couplings is not yet available, we cannot be completely sure that
the relevant couplings are distinct, or indeed that they are non-vanishing, but at least we
can inspect if there are obvious symmetries relating them.

The unextended tensor product $(2,4,22,22)$ has
64 discrete symmetries: five separate charge conjugations of the factors (including the
NSR space-time factor) and the interchange of the two identical $k=22$ minimal models.
To get space-time supersymmetry this tensor product is extended with the
product of the simple current Ramond ground states of each factor. These Ramond
ground states are not invariant under charge conjugation. Therefore this extension
breaks the discrete symmetries to ${\bf Z}_2 \times {\bf Z}_2$, the combined charge
conjugation of all five factors and the permutation of the two identical factors. The
combined charge conjugation also acts non-trivially on the Ramond ground states
in each factor, but the result is the charge conjugate of the space-time 
supersymmetry generator, which is in the chiral algebra. The combined conjugation
is in fact the charge conjugation symmetry of the extended CFT.
It turns out that 
only a ${\bf Z}_2 $  subgroup of ${\bf Z}_2 \times {\bf Z}_2$ acts non-trivially on
the simple currents of the extended CFT: the permutation of the $k=22$ factors
acts in the same way as charge conjugation. The action of these symmetries on
the complete set of primary fields is more complicated.  It is easy to see that the
permutation acts differently than charge conjugation. In general  the primary fields of the
extended CFT are labelled as $i=[((x),(i_1),\ldots,(i_4)),\Psi]$ (see Appendix A).
The action of the permutation is to interchange
$i_3$ and $i_4$, but in cases with a non-trivial degeneracy it is not {\it a priori} clear
which of the degenerate states is the image of the map. This can be resolved by
examining the fusion rules, which should be invariant under the permutation:
\beqa
 N_{ijk}= N_{\pi(i)\pi(j)\pi(k)} \ ,
\eeqa
where $N$ is  a  fusion coefficient and $\pi$ a permutation or other automorphism.
 In general
there may be more than one way to resolve these ambiguities, resulting in additional
automorphism of  the CFT. The standard example of this situation is the extension
of the affine algebra $A_1$ level 4 by the simple current. The resulting CFT has an
outer automorphism, non-trivial charge conjugation, that has no counterpart in $A_1$ 
level 4.

As mentioned above, the ${\bf Z}_2$
permutation symmetry is respected by the MIPF and the orientifold, and since charge
conjugation acts in the same way on the simple currents as the permutation, charge
conjugation is respected as well. 

In this way we end up with (at least)  a surviving ${\bf Z}_2 \times {\bf Z}_2$
discrete symmetry acting on the boundary labels, or a larger discrete symmetry
if that symmetry is extended by the action on the degeneracy labels of the extension.
The foregoing story repeats itself for the action on the boundary labels. The boundary
labels are given in terms of the CFT labels plus a second degeneracy label, the one
indicated  by the second entry in the square brackets in column 2 of tables (\ref{tbl:SMBoundaries}) and 
(\ref{tbl:InstantonBranes}). Once again one has to determine not only how 
a symmetry acts on the first entry (this is just the action of the symmetry in the
extended CFT, respecting its fusion rules), but also how it acts on the degeneracy labels.
In this case the precise action can be determined from the invariance of the
annulus coefficients
\beqa
 A^i_{ab}=A^{\pi(i)}_{\hat \pi(a)\hat \pi(b)} \ ,
\eeqa
where $\pi$ is the action on the primaries of the extended CFT (as determined above)
and $\hat \pi$ is the action on the boundary labels induced by $\pi$.

Since the orientifold choice is non-trivial, boundary charge conjugation does not
coincide with CFT charge conjugation. Indeed, the eight instanton boundary states
are invariant under boundary charge conjugation (which they must be in order
to produce a ``real" $Sp(2)$-type instanton).  However, just as permutations,
CFT charge conjugation may induce a non-trivial discrete symmetry on the
boundary states. 

In addition to these ``outer automorphisms" there is the notion of boundary
simple currents, introduced in the appendix of \cite{schell1}. These may be thought
of as remnants of the original simple currents, and imply relations
between annulus amplitudes of the form
\beqa
A^i_{ab'}=A^i_{Ja (Jb)'}
\eeqa

All of the aforementioned symmetries might relate instanton couplings, and hence
threaten their numerical independence.
However, in order to do that they have to be symmetries of the full
standard model/instanton configuration, not just relate some of the eight instantons
to each other. 
It is easy to see that the permutation of the $k=22$ factors changes the standard model
brane configuration. Consider brane {\it c}: it turns out that under permutation boundary
state 629
it is mapped to boundary state
544 or 545 (depending on the action on the degeneracy label), which in any case
is distinct. Hence even if the instanton boundaries 414 and 456 resp. 418 and 502
are mapped to each other by boundary permutation, at the same time
the standard model configuration is mapped to a distinct one.

This means that we may expect at least four distinct couplings, which should be
sufficient. It is of course possible to work out the discrete symmetries exactly,
but in view of this argument this would not yield any additional insight.

We do know the exact boundary orbits. The orbit of instanton label 414 is\\
$(414,415,416,417)$, so that instantons 414 and 417 are related. But the
orbit of brane ${\it c}$  under the same action is $(629,628,626,627)$. Hence
the action that relates 414 and 417 maps 629 to 627. In fact all four
standard model boundaries {\bf a,b,c,d} are mapped to different ones.
This implies that instantons 414 and 417 may produce different
couplings as well, so that all eight instantons may contribute in a different way.

These distinctions concern the disk correlators $d_a^{(r)}$ in (\ref{InstSum}).
The factors ${\rm exp}(-{{\rm Re~}} U_r)$  {\it will} be related by discrete symmetries,
and it seems reasonable to expect them to be identical for instantons 414, 417,
456  and 459, which is indeed correct.
However there is no reason to expect the other four instantons
to have the same suppression factor, and indeed they do not.

Note that these symmetries imply the existence of a much larger set
of standard model configurations than the 32 discussed here. However,
as mentioned before, the 32 models considered here display all 
possible distinct spectra.

 \newcommand\wb{\,\linebreak[0]} \def\wB {$\,$\wb}
 \newcommand\Bi[1]    {\bibitem{#1}}
 \newcommand\J[5]   {{\sl #5}, {#1} {#2} ({#3}) {#4} }
 \newcommand\Prep[2]  {{\sl #2}, preprint {#1}}
 \def\jf    {J.\ Fuchs}
 \def\adma  {Adv.\wb Math.}
 \def\anop  {Ann.\wb Phys.}
 \def\aspm  {Adv.\wb Stu\-dies\wB in\wB Pure\wB Math.}
 \def\atmp  {Adv.\wb Theor.\wb Math.\wb Phys.}
 \def\comp  {Com\-mun.\wb Math.\wb Phys.}
 \def\ijmp  {Int.\wb J.\wb Mod.\wb Phys.\ A}
 \def\jhep  {J.\wb High\wB Energy\wB Phys.}
 \def\mpla  {Mod.\wb Phys.\wb Lett.\ A}
 \def\nuci  {Nuovo\wB Cim.}
 \def\nupb  {Nucl.\wb Phys.\ B}
 \def\phlb  {Phys.\wb Lett.\ B}
 \def\phrl  {Phys.\wb Rev.\wb Lett.}
 \def\NH     {{North Holland Publishing Company}}
 \def\SV     {{Sprin\-ger Ver\-lag}}
 \def\WS     {{World Scientific}}
 \def\Ad     {{Amsterdam}}
 \def\Be     {{Berlin}}
 \def\Si     {{Singapore}}

\end{document}